\documentclass[reqno,10pt,a4paper,dvips]{article}

\usepackage{amssymb}
\usepackage{amsmath}
\usepackage{amsthm}
\usepackage{cite}
\usepackage{psfrag}
\usepackage{array}
\usepackage{setspace}
\usepackage{geometry}
\usepackage{color}
\usepackage[dvips]{graphicx}
\usepackage{amscd}
\usepackage{enumitem}
\usepackage{verbatim}
\usepackage{epsfig}

\newcolumntype{C}{>{$}c<{$}} 
\newcolumntype{R}{>{$}r<{$}} 

\newcommand{\Dir} {\mathrm{D}}

\newcommand{\fvanish} {H}
\newcommand{\length}[1] {{L_{#1}}}
\newcommand{\supp}[1] {{\mathrm{supp}(#1)}}
\newcommand{\inv} {\mathrm{inv.}}

\newcommand{\set}[1]{\left\{ {#1} \right\}}

\newcommand{\wtil} {\widetilde}

\newcommand{\half} {\frac{1}{2}}
\newcommand{\const} {\mathrm{const.}}

\newcommand{\im} {\Im \mathrm{m } \, }
\newcommand{\re} {\Re \mathrm{e } \, }
\newcommand{\imof}[1] {\Im \textrm{m} \left( #1 \right)}

\newcommand{\cc} {\overline}

\newcommand{\resatof}[2] {\mathrm{Res}_{#1} \left( #2 \right)}

\newcommand{\cl} {\overline}
\newcommand{\bdry} {\partial}

\newcommand{\EX} {\mathsf{E}}

\newcommand{\stopping} {\sigma}
\newcommand{\stoppingBis} {{\sigma'}}

\newcommand{\dist}{\mathrm{dist}}
\newcommand{\rapdec}{\mathcal{S}}
\newcommand{\tempdist}{{\rapdec'}}

\providecommand{\bra} {\big<}
\providecommand{\ket} {\big>}
\providecommand{\der}[1]{\frac{\ud}{\ud {#1}}}

\providecommand{\pder}[1]{\partial_{{#1}}}
\providecommand{\pderS}[1]{\partial_{{#1 #1}}}

\providecommand{\nder}{{\partial_{n}}}
\providecommand{\tder}{{\partial_{\mathbf \tau}}}
\providecommand{\lapl}{\triangle}
\providecommand{\Ord}[1]{\mathcal{O} \left( {#1} \right)}
\providecommand{\harmmeas}{\omega}

\newcommand{\hull} {K}

\newcommand{\lochcap} {\mathrm{lhcap}}

\newcommand{\confmap} {\phi} 
\newcommand{\sign} {\mathrm{sign}}

\newcommand{\domain}{\Omega}

\newcommand{\eps} {\varepsilon}

\newcommand{\bH} {\mathbb{H}}  
\newcommand{\bC} {\mathbb{C}}  
\newcommand{\bR} {\mathbb{R}}  
\newcommand{\bZ} {\mathbb{Z}}  
\newcommand{\bD} {\mathbb{D}}  
\newcommand{\bS} {\mathbb{S}}  
\newcommand{\bA} {\mathbb{A}}  
\newcommand{\sF} {\mathcal{F}} %
\newcommand{\ud} {\mathrm{d}}
\newcommand{\ii}{\mathfrak{i}}

\newcommand{\eqnref}[1]{Equation~(\ref{#1})}
\newcommand{\eqnDref}[2]{Equations~(\ref{#1}) and (\ref{#2})}

\newcommand{\secref}[1]{Section~\ref{#1}}
\newcommand{\secDref}[2]{Sections~\ref{#1} and \ref{#2}}

\newcommand{\thmref}[1]{Theorem~\ref{#1}}

\newcommand{\propref}[1]{Proposition~\ref{#1}}
\newcommand{\propDref}[2]{Propositions~\ref{#1} and \ref{#2}}

\newtheorem{theorem}{Theorem} 

\newtheorem{proposition}[theorem]{Proposition}

\newtheorem{remark}{Remark}

\numberwithin{equation}{section}

\begin{document}

\title{Hadamard's formula and couplings of SLEs with free field}


\date{}

\maketitle

\begin{center}
Konstantin Izyurov\footnote{\texttt{Konstantin.Izyurov@unige.ch}}
and Kalle Kyt\"ol\"a\footnote{\texttt{Kalle.Kytola@unige.ch}} \\
Universit\'e de Gen\`eve, Section de Math\'ematiques
\end{center}

\begin{abstract}
The relation between level lines of Gaussian free fields (GFF)
and SLE${}_4$-type curves was discovered by O.~Schramm and
S.~Sheffield. A weak interpretation of this relation is the existence
of a coupling of the
GFF and a random curve, in which the curve behaves like a level line
of the field. In the present paper we study these couplings for the
free field with different boundary conditions. We provide a unified
way to determine the law of the curve (i.e. to compute the driving
process of the Loewner chain) given boundary conditions of the field,
and to prove existence of the coupling. The proof is reduced to the
verification of two simple properties of the mean and covariance of
the field, which always relies on Hadamard's formula and properties of
harmonic functions.

Examples include combinations of Dirichlet, Neumann and Riemann-Hilbert
boundary conditions. In doubly connected domains, the standard
annulus SLE${}_4$ is coupled with a compactified GFF obeying Neumann
boundary conditions on the inner boundary. We also consider variants
of annulus SLE coupled with free fields having other natural boundary
conditions. These include 
boundary conditions leading to curves connecting two points on different
boundary components with prescribed winding
as well as those recently proposed by
C.~Hagendorf, M.~Bauer and D.~Bernard.

\end{abstract}


%
%

\section{Introduction}
\label{sec: intro}

The topic of conformally invariant random processes in two dimensions has
received a lot of attention during the past decade.
Recent developments have enabled a
probabilistic approach to problems traditionally studied in theoretical
physics by means of conformal field theory. 

Two fundamental examples of random conformally invariant objects are
Schramm-Loewner evolutions (SLE) and Gaussian free fields (GFF).
Schramm-Loewner evolutions are random fractal curves
described by growth processes encoded in Loewner chains.
Their most important characteristics are captured by one parameter,
a positive real number $\kappa$, but still in different setups one needs different
variants of SLE${}_\kappa$ as we will again see in this article.
The Gaussian free field is a statistical model that fits naturally both in
the setup of conformal field
theory and in that of probability theory: it is essentially the simplest
Euclidean quantum field theory, which describes the free massless boson,
but it also admits an easy interpretation as a random generalized function.

Informally speaking, the Gaussian free field $\Phi$ in a planar domain $\domain$
is a collection of Gaussian random variables indexed by the points of the domain,
$\Phi = \big( \Phi(z)\big)_{z \in \domain}$, such that:
\begin{itemize}
\item The mean $\EX \left[ \Phi(z) \right] = M(z)$ is a harmonic function.
\item The covariance $\EX \left[ \big( \Phi(z_1) - M(z_1) \big) 
\big( \Phi(z_2) - M(z_2) \big) \right] = C(z_1, z_2)$
is a Green's function in $\domain$.
\end{itemize}
To obtain an unambiguous definition of the GFF one has to specify
which harmonic function to choose, and what is meant by the Green's
function. We will usually specify $M$ by its boundary conditions. The
Green's functions will be solutions to
$-\lapl G(\cdot,z_2) = \delta_{z_2}(\cdot)$ with prescribed boundary
conditions. From the definition one immediately sees that GFF will posses
conformal invariance properties --- indeed harmonic functions and Green's
functions are simply transported by conformal maps. If
$\confmap : \domain \rightarrow \domain'$ is a conformal map
and $\Phi$ is a GFF in $\domain'$, then $\Phi\circ \confmap$ is a GFF in
$\domain$, boundary conditions in $\domain$ being the pullback of those
in $\domain'$.
We will mostly deal with boundary conditions that transform nicely under
conformal maps.

Note that $\Phi$ being Gaussian, the law is indeed
determined by its mean and covariance. Due to the blowup of the
covariance as $|z_1-z_2| \rightarrow 0$, however, the field $\Phi$
is not a random function but rather a random distribution
(a generalized function).
We postpone a formal definition of GFF to \secref{sec: coupling}.

A typical example of how the mean $M$ and covariance $C$ are specified
appears in the works of Schramm and Sheffield
\cite{SS-harmonic_explorer, SS-contour_lines} which first established
a relation between the Gaussian free field and Schramm-Loewner evolutions.
In a simply connected domain $\domain$ with boundary $\bdry \domain$
divided to two complementary arcs $l_1$ and $l_2$ one defines $M$ and $C$ by
\begin{align} \label{eq: SS example}
\left\{ \begin{array}{rll}
\lapl M(z) & = \; 0 \quad & \textrm{ for $z \in \domain$}\\
M(z) & = \; +\lambda \quad & \textrm{ for $z \in l_1$} \\
M(z) & = \; - \lambda \quad & \textrm{ for $z \in l_2$}
\end{array} \right.
\qquad \textrm{ and } \qquad
\left\{ \begin{array}{rll}
\lapl_z C(z,z_2) & = \; -\delta_{z_2}(z) \quad & \textrm{ for $z \in \domain$}\\
C(z,z_2) & = \; 0 \quad & \textrm{ for $z \in \bdry \domain$.}
\end{array} \right. 
\end{align}
Schramm and Sheffield showed that chordal SLE${}_4$ describes the scaling
limit of the zero level lines of a discrete Gaussian free field with the
above boundary conditions, when the parameter $\lambda$ has the particular
value $\lambda = \sqrt{\pi/8}$.
In particular, the free field is naturally coupled with a chordal
SLE${}_4$, and in the scaling limit the level lines of discrete GFF
become discontinuity lines of the GFF of jump $2 \lambda$. 

We will be interested in couplings of different variants of GFF with
random growth proceces of SLE type.
A variant of GFF is a rule associating
to any domain $\domain$ with $n+1$ marked points
$x, x_1, x_2, \dots, x_n \in \bdry \domain$
a free field $\Phi_{\Omega;x, x_1,\dots,x_n}$ --- for instance, in the above
example (\ref{eq: SS example}) the marked points are the two endpoints of
the boundary arcs. Take a domain $(\domain;x,x_1,x_2,\dots)$ and suppose
we have a random curve $\gamma \subset \Omega$ growing from $x$.
The main property we require of the coupling is:
\begin{quotation}
\label{quote: coupling property}
Conditionally on the random curve $\gamma \subset \domain$ starting from
$x \in \bdry \domain$, the law of
the free field $\Phi_{(\domain;x,x_1,\ldots,x_n)}$ is the same as that of
the free field $\Phi_{(\wtil{\domain};\tilde{x},x_1,\ldots,x_n)}$ in the
domain $\wtil{\domain} = \domain \setminus \gamma$, where
$\tilde{x} \in \bdry \wtil{\domain}$ is the tip of the curve $\gamma$.
\end{quotation}
This property also immediately suggests a constructive way of producing
the coupling given the random curve and the laws of the free fields in
different domains:
\begin{quotation}
\label{quote: sampling}
Sample the random curve $\gamma$, and then sample independently the free
field $\Phi_{(\wtil{\domain}; \tilde{x}, x_1, \ldots, x_n)}$ in the slitted
domain $\wtil{\domain} = \domain \setminus \gamma$. The law of the resulting
field is the same as the free field $\Phi_{(\domain; x, x_1, \ldots, x_n)}$
in the original domain.
\end{quotation}
The motivation for imposing these properties of the coupling is
the example of Schramm \& Sheffield, in which the discontinuity line
of the free field satisfies them. The present article exhibits numerous
variations of that basic example.

\bigskip

The article is organized as follows.
\secDref{sec: Loewner chains}{sec: SLE}
recall necessary background on Loewner chains and SLE in simply and
multiply connected domains. \secref{sec: couplings of SLE and GFF}
is devoted to the general setup for establishing couplings of SLEs
and free fields.
\secref{sec: basic equations} writes the two basic conditions that we
will verify in each case to prove the existence of couplings, and 
\secref{sec: SSExample} concretely illustrates these conditions in
the simplest example case (\ref{eq: SS example}) of Schramm \& Sheffield.
We define free field in \secref{sec: coupling} and show that the two
basic conditions imply a weak form of coupling. Next, in
\secref{sec: Hadamard} we recall and
prove in a setup appropriate for the present purpose the Hadamard's
formula, whose variants are crucial to the verification of the basic
conditions in all cases.

The concrete examples are divided to two sections, treating simply connected
domains and doubly connected domains separately.
\secref{sec: simply connected} presents free fields with
different boundary conditions in simply connected domains. The examples here
include coupling of the dipolar SLE${}_4$ with GFF having combined
jump-Dirichlet and Neumann boundary conditions, and the coupling of
SLE${}_4(\rho)$ with GFF having combined jump-Dirichlet and Riemann-Hilbert
boundary conditions. We also show in the presence of more complicated
combinations of boundary conditions how the coupling determines
the law of the curve, i.e. how to compute the Loewner driving process.
\secref{sec: doubly connected} treats examples in doubly connected
domains. We warm up with a simple case of a punctured disc, giving a
short proof that the radial SLE${}_\kappa$ is coupled with a compactified free
field with jump-Dirichlet boundary conditions as stated in
\cite{Dubedat-SLE_and_free_field}.
In an annulus with jump-Dirichlet boundary conditions on one boundary
component and Neumann boundary conditions on the other, we show that the compactified
free field is coupled with the standard annulus SLE${}_4$ introduced in
\cite{BB-zig_zag, Zhan-SLE_in_doubly_connected_domains}.
We also review the SLE${}_4$ variants proposed in
\cite{HBB-free_field_in_annulus} on grounds of free field partition functions,
and show that they indeed admit couplings with the non-compactified free fields
with corresponding boundary conditions. Another new example consists in
imposing jump-Dirichlet boundary conditions on both boundary components for
a compactified free field, leading to a curve with prescribed winding.
In \secref{sec: Dirichlet general kappa} we show that the cases with
Dirichlet boundary conditions admit generalizations to $\kappa \neq 4$.

Appendix \ref{sec: commutation} explains why extensions
at $\kappa \neq 4$ don't work with all boundary conditions, and
Appendix \ref{sec: Loewner lemma} contains the proof of a property
of Loewner chains we need in conjunction with the general Hadamard's
formulas

\bigskip

\subsubsection*{Relation to other work}
We note that the relation of the free field and SLE has already been
explored beyond the basic example of Schramm and Sheffield.
One research direction has been establishing the coupling in a strong sense.
Note that in this article we
content ourselves with the weak form of the coupling
described above, and we only consider restriction
of the free field to subdomains almost surely untouched by the curve.
Dub\'edat has however given a procedure to extend couplings from subdomains
to the full domain, and shown the strong interpretation of the coupling
in which the random curve is a deterministic function of the free field
configuration \cite{Dubedat-SLE_and_free_field}. 

The effect of the boundary conditions of the free
field on the law of the curve is another important generalization of
the basic example,
and this is the direction we systematically pursue also in the present
article. Earlier work in this direction concerns especially
the appropriate SLE variants when one allows several jumps in
the Dirichlet boundary conditions, discussed in some cases already in
\cite{SS-contour_lines, Cardy-SLE_kappa_rho} and developed in
more generality in \cite{Dubedat-SLE_and_free_field}.
Recently, Hagendorf \& Bauer \& Bernard \cite{HBB-free_field_in_annulus}
proposed natural SLE variants in annulus based on computations of free
field partition functions with combined jump-Dirichlet and Neumann
boundary conditions. Our examples cover also these cases explicitly.

It is also worth noting that Schramm \& Sheffield themselves indicated
how their coupling can be extended to chordal SLE${}_\kappa$ with
$\kappa \neq 4$ by modifying the conformal transformation property
of the field in the manner dictated by the Coulomb gas formalism of
conformal field theory. In our examples which involve piecewise Dirichlet
boundary conditions we show how to treat $\kappa \neq 4$, and we give
a non-commutation argument explaining why one is constrained to $\kappa=4$
in the presence of other boundary conditions.

Generalizations to massive free fields have been treated in
\cite{MS-massive_SLEs_ICMP, BBC-near_critical}. Many aspects of SLE${}_4$
related conformal field theories are considered in the forthcoming articles
\cite{MZ-free_fields, KM-free_fields}.

\subsection{Growth processes and Loewner evolutions}
\label{sec: Loewner chains}
The Loewner evolution is a way of describing growth processes,
curves in particular,
in terms of conformal maps.
In the case when $\domain$ is a simply-connected domain with
analytic boundary, a setup convenient for our purposes is as follows.
To each point of the boundary $x \in \bdry \domain$
we associate a Loewner vector field 
$V_x(z) \pder{z}$, satisfying the following properties
\begin{itemize}
\item $V_x(z)$ is analytic inside the domain $\domain$ and up to the boundary
apart from the point $x$;
\item for $z \in \bdry \domain \setminus \set{x}$ the vector field $V_x(z) \pder{z}$
is tangential to the boundary;
\item $V_x(z)$ has a simple pole at $x$ with 
$\resatof{x}{V_x(z)} = 2 \, \tau_x^2$, where $\tau_x$ is a unit tangent to $\bdry \domain$
at $x$;
\item $V_x(z)$ is bounded apart from the neighborhood of $x$.
\end{itemize}

Given a continuous function $t \mapsto X_t \in \bdry \domain$ called the
driving process, the Loewner's differential equation is
\begin{align*}
\label{eq: Loewner}
\der{t} g_t(z) \; = \; V_{X_t}(g_t(z)), \qquad g_0(z) = z
\tag{Loe}
\end{align*}
where the initial condition is a point of the domain, $z \in \domain$.
For all $t\geq0$ we let $\hull_t \subset \domain$ be the set of
points $z$ for which the solution fails to exist up to time $t$.
The hulls $(\hull_t)_{t \geq 0}$ form a growth process,
$\hull_{t_1} \subset \hull_{t_2}$ for $t_1 < t_2$. The solution
$(g_t)_{t \geq 0}$ is called a Loewner chain for the growth process.

Familiar examples in the half-plane, disc and strip are respectively
\begin{align*}
& \textrm{{\bf Domain}} & & \textrm{{\bf Vector fields}}
    & & \textrm{{\bf Flow}} &
\vspace{.2cm} \\
\label{eq: Loewner H}
    & \bH = \set{\im z > 0} \qquad & &
    V_x(z) = \frac{2}{z-x} \qquad & & \der{t} g_t(z) = \frac{2}{g_t(z)-X_t}
\tag{Loe-$\bH$} 
\\
\label{eq: Loewner D}
& \bD = \set{|z|<1} \quad & &
    V_x(z) = -z \frac{z+x}{z-x} \quad & &
    \der{t} g_t(z) = g_t(z) \frac{X_t+g_t(z)}{X_t-g_t(z)}
\tag{Loe-$\bD$} 
\\
\label{eq: Loewner S}
& \bS = \set{0 < \im z < \pi} \quad & &
    V_x(z) = \coth \left( \frac{z-x}{2} \right) \quad & &
    \der{t} g_t(z) = \coth \left( \frac{g_t(z)-X_t}{2} \right).
\tag{Loe-$\bS$}
\end{align*}
The first flow in $\bH$ fixes the point $\infty$, the second in $\bD$ fixes $0$,
and the third in $\bS$ fixes both $\pm \infty$. These properties make the chosen
flows convenient, but we remark that the choices are by no means unique.
In particular it is worth noting that a growth process can be described
by several different Loewner chains.
In what follows we assume that $V_x(z)$ depends sufficiently nicely on $x$, as is
the case with the three examples.

The following proposition is standard, and in concrete examples we only use it
with the three vector fields listed above.
\begin{proposition} \label{prop: hloc}
For all $t>0$, $\domain \setminus \hull_t$ is simply-connected, and
$z \mapsto g_t(z)$ is a conformal map from
$\domain_t := \domain \setminus \hull_t$ to $\domain$.
Moreover, $\pder{t} \lochcap_{X_0} (\hull_t)|_{t=0} = 2$, where
$\lochcap$ is the local half-plane capacity.
\end{proposition}

The local half-plane capacity of $\hull_t$ is informally defined as follows: if the
boundary near the point $X_0$ is a straight line, translate and rotate the domain
so that it would actually become a part of $\bR$ and $\hull_t$ would become a subset of
$\bH$, then take the half plane capacity.
If the boundary is not a straight line, use a conformal map $f$ to $\mathbb{H}$ such
that $|f'(x_0)|=1$. Roughly speaking, the last statement of the Proposition
means that for small values of $t$, the size of the hull doesn't depend much on
global structure of the domain and the evolution --- to first order it is completely
determined by the residue of $V_x(z)$ (which in turn is fixed by our conventions
for Loewner vector fields). We postpone the formal definition along with the proof
of the Proposition to Appendix \ref{sec: Loewner lemma}.

Note that the map $g_t$ maps the tip of the growing hull $\hull_t$
to the point $X_t$.
The notion of the tip is intuitive if the hulls are
growing curves, $\hull_t = \gamma[0,t]$.
It is, however, always well-defined, since the Loewner
chain satisfies the \textit{local growth property}:
$\lim_{\eps \searrow 0} (\cl{\hull_{t+\eps} \setminus \hull_t})$
is always a boundary point of $\domain \setminus \hull_t$
(more precisely, a prime end). We call this point the tip of the hull
and denote it by $\tilde{x}(t)$.

\subsubsection*{Loewner chains in doubly connected domains}
\label{sec: LoewnerNSC}
For multiply connected domains the Loewner flow $V_x(z)$ cannot be tangential
to the boundary on all boundary components --- once we start growth, the
conformal moduli of the domain change, and $z \mapsto g_t(z)$ cannot be a
map from $\domain \setminus \hull_t$ onto $\domain$ anymore.
Hence, instead of one domain, we fix a family
of representatives of conformal equivalence
classes, and $g_t$ maps to one of these. In doubly connected case,
a natural family is provided by the annuli
$\bA_r=\{ z \in \bC \; : \; e^{-r}<|z|<1\}$, $r>0$,
with the unit circle as their common boundary
component. For the Loewner flow to preserve this family,
the radial component of the vector field
should be constant on the inner boundary circle.
This is equivalent to the condition
$\re \left( V^r_x(z)/z \right) = C$ on $|z|=e^{-r}$.
On the outer component of the boundary, we want
$V_x^r(z) \pder{z}$ to be tangential to the boundary, meaning
$\re \left(V^r_x(z)/z \right) = 0$ for $|z|=1$, $z \neq x$.
For any value of the constant $C$, there
exists a unique harmonic function with such boundary conditions
and desired singularity at $x$, but only at one value of $C$ the harmonic
conjugate becomes a single-valued function. Namely, there exists a unique
function $S^r_x(z)$ (Schwarz kernel) satisfying the following properties:
\begin{itemize}
\item $S_x^r(z)$ is analytic in the annulus $\bA_r$;
\item $\re S_x^r(z)=\delta_x(z)$ on the outer boundary $\set{|z|=1}$;
\item $\re S_x^r(z)=\frac{1}{2\pi}$ on the inner boundary $\set{|z|=e^{-r}}$.
\end{itemize}
There is a complicated explicit expression for $S_x^r(z)$
\cite{BB-zig_zag, Zhan-SLE_in_doubly_connected_domains},
but we will not need it.

We define Loewner vector fields as $V^r_x(z)=2\pi z S_x^r(z)$.
With this choice the modulus $r$ decreases at
unit speed under the flow analogous to (\ref{eq: Loewner}): if
$\domain = \bA_p$, then $g_t(\domain \setminus \hull_t) = \bA_{p-t}$.
The modulus therefore directly serves as a time parametrization of the
Loewner chain.
\begin{align*}
& \textrm{{\bf Domains}} & & \textrm{{\bf Vector fields}} & &
\textrm{{\bf Flow}} 
\\
\label{eq: Loewner A}
& \bA_r = \set{ e^{-r}<|z|<1} \qquad & & 
    V^r_x(z) = 2\pi \, z \, S_x^r(z) \qquad & &
    \der{t} g_t(z) \; = \; 2 \pi \, z \, S^{p-t}_{X_t}(z) .
\tag{Loe-$\bA$}
\end{align*}
The analogue of \propref{prop: hloc} remains valid for this Loewner
chain on the time interval $t \in [0,p)$.

\subsection{Schramm-Loewner evolutions}
\label{sec: SLE}
Stochastic Loewner evolutions (or Schramm-Loewner evolutions, SLE)
are random growth processes defined via a Loewner chain with random
driving process. The random driving process is chosen
so that the growth process satisfies two fundamental properties:
\emph{conformal invariance} and \emph{domain Markov property} ---
the reader is referred to one of
the many excellent introductions to SLE for details, e.g.
\cite{Werner-random_planar_curves, Lawler-conformally_invariant_processes,
BB-2d_growth_processes}.
In particular the driving process will always be chosen to be a
semimartingale (living on the boundary of the domain) whose quadratic
variation grows at constant speed $\kappa>0$, indicated by a subscript
SLE${}_\kappa$. In the following well known examples the driving process
is simply a Brownian motion on $\bdry \domain$ with the appropriate
speed --- here and in the sequel $(B_t)_{t\geq0}$ stands for a
standard Brownian motion on $\bR$:
\begin{itemize}
\item {\bf Chordal SLE${}_\kappa$ in $\bH$ from $0$ to $\infty$:} \\
The Loewner chain is (\ref{eq: Loewner H}) with the driving process
$X_t = \sqrt{\kappa} \,B_t$.
\item {\bf Radial SLE${}_\kappa$ in $\bD$ from $1$ to $0$:} \\
The Loewner chain is (\ref{eq: Loewner D}) with the driving process
$X_t = \exp (\ii \sqrt{\kappa} \, B_t)$.
\item {\bf Dipolar SLE${}_\kappa$ in $\bS$ from $0$ to $\bR + \ii \pi$:} \\
The Loewner chain is (\ref{eq: Loewner S}) with the driving process
$X_t = \sqrt{\kappa} \, B_t$. Note also that this is a special case of the
example of SLE${}_\kappa(\rho)$ in $\bS$ below,
with $\rho = \frac{\kappa-6}{2}$.
\end{itemize}

In other examples the driving process $X$ may have a drift. For instance,
if the domain has marked points $x, x_1,x_2,\dots, x_n$ on the boundary,
then the slitted domain $(\Omega_t,\tilde{x}(t),x_1,\dots,x_n)$ is in general
not conformally equivalent to $\Omega,x_1,x_2,\dots, x_n$. 
The drift term of the It\^o diffusion may therefore depend on
conformal moduli of this configuration, as in the first of
the following two examples:
\begin{itemize}
\item {\bf SLE${}_{\kappa}(\overline{\rho})$ in $\bH$ started from $0$:} \\
Here $\overline{\rho}=(\rho_1, \rho_2,\dots,\rho_n)$ is an $n$-tuple of real
parameters. The marked points other than $x=0$ are
$x_1, x_2, \dots, x_n \in \bR$ on the boundary.
The Loewner chain is (\ref{eq: Loewner H}) with driving process obeying the
It\^o diffusion
$\ud X_t = \sqrt{\kappa} \, \ud B_t
+ \sum_j \frac{\rho_j}{X_t-g_t(x_j)} \, \ud t$, with $X_0=x=0$.
\item {\bf SLE${}_\kappa(\rho)$ in $\bS$ started from $0$:} \\
In the above example if $n=1$ it is convenient to perform a coordinate change
from $\bH$ to $\bS$ sending $0 \mapsto 0$, $x_1 \mapsto \pm \infty$ and
$\infty \mapsto \mp \infty$, see e.g. \cite{Kytola-SLE_kappa_rho}.
The resulting growth process is described,
up to a time reparametrization, by a Loewner chain (\ref{eq: Loewner S})
with driving process
$X_t = \sqrt{\kappa} \, B_t \mp ( \rho + \frac{6-\kappa}{2} ) \, t$.
\end{itemize} 

The simplest example of SLE${}_\kappa$ in doubly connected domains is the
following, proposed independently in
\cite{BB-zig_zag, Zhan-SLE_in_doubly_connected_domains}:
\begin{itemize}
\item {\bf Standard annulus SLE${}_\kappa$ in $\bA_p$ started from $1$:} \\
The Loewner chain is (\ref{eq: Loewner A})
with driving process $X_t = \exp(\ii \sqrt{\kappa} \, B_t)$.
\end{itemize}

There are, of course, more variants of SLE${}_\kappa$.
We will find natural free fields admitting coupling with each of the
above examples --- and when boundary conditions of the free field are more
complicated, we find other variants.

\section{Couplings of SLEs and Gaussian Free Fields}
\label{sec: couplings of SLE and GFF}

\subsection{Basic equations}
\label{sec: basic equations}
Recall that we're interested in Gaussian free fields coupled with random
curves or growth processes in the way described in the introduction.
Suppose we have a rule associating to each domain
$\domain$ (with marked points) a free field $\Phi_\domain$, determined by a harmonic
function $M_\domain : \domain \rightarrow \bR$ and a Green's function
$C_{\domain}: \domain \times \domain \setminus \set{z_1 = z_2} \rightarrow \bR$.
Consider a random growth process of hulls $(\hull_t)_{t \in [0,\stopping]}$ in a domain
$\domain_0$ and let $\domain_t = \domain_0 \setminus \hull_t$. Now construct a
field $\wtil{\Phi}$ by first sampling the final
random hull $\hull_\stopping$, and then on the remaining random domain $\domain_\stopping$
sampling an independent free field with the law of $\Phi_{\domain_\stopping}$.
Does the law of $\wtil{\Phi}$ coincide with the law of $\Phi_{\domain_0}$,
at least on a subset of $\domain_0$ that is almost surely untouched by $K_\stopping$?

A necessary condition for the field $\wtil{\Phi}$ to have the same law as
$\Phi_{\domain_0}$ is that the mean and covariance coincide, which can be written as
\begin{align}
\label{eq: M-weak}
M_{\domain_0}(z) \; \overset{?}{=} \; & \EX \left[ M_{\domain_\stopping}(z) \right]
\\
\label{eq: C-weak}
C_{\domain_0}(z_1,z_2) + M_{\domain_0} (z_1) M_{\domain_0}(z_2)
\; \overset{?}{=} \; & \EX \left[
C_{\domain_\stopping}(z_1,z_2) + M_{\domain_\stopping}(z_1) M_{\domain_\stopping}(z_2) \right]
\end{align}
for all $z$ in the domain where $\wtil{\Phi}$ is defined. The expected values
here refer to averages over the random hull $\hull_\stopping$.

If we knew a priori that $\wtil{\Phi}$ is Gaussian, then the conditions
(\ref{eq: M-weak}) and (\ref{eq: C-weak}) would imply the desired
coincidence of laws of $\wtil{\Phi}$ and
$\Phi_{\domain_0}$, since the two Gaussian variables would have equal means and
covariances. We will actually impose the following stronger conditions, from
which the coincidence of laws will follow, as will be proven in
\secref{sec: coupling}. We require that
\begin{align*}
\label{M-eq}
    M_t(z) \; := \; M_{\domain_t}(z)
    \; \textrm{ are uniformly bounded continuous martingales}
\tag{M-cond} \\
\label{C-eq}
    \textrm{such that } \;
    \bra M(z_1), M(z_2) \ket_t \; = \;
    C_{\domain_0}(z_1, z_2) - C_{\domain_t}(z_1,z_2).
\tag{C-cond}
\end{align*}
Here $\bra \cdot, \cdot \ket_t$ denotes the quadratic cross variation ---
the second condition is therefore equivalent to
$t \mapsto C_{\domain_t} (z_1,z_2)$ being a process of finite variation
such that $M_t(z_1)M_t(z_2)+C_{\domain_t} (z_1,z_2)$ is martingale.
Note that by optional stopping theorem for the martingales
$M_t(z)$ and $M_t(z_1) M_t(z_2) + C_{\domain_t}(z_1,z_2)$ at time
$\stopping$ the above conditions
indeed guarantee (\ref{eq: M-weak}) and (\ref{eq: C-weak}).

In practise verifying the two basic conditions becomes rather explicit. We
mostly deal with strictly conformally invariant boundary conditions
in the following sense. Consider simply connected domains $\domain$ with
$n+1$ marked points $x,x_1,x_2,\ldots,x_n \in \bdry \domain$, and associate
to them harmonic functions $M_{(\domain;x,x_1,\ldots,x_n)}$ defined on
$\domain$ and Green's functions $C_{(\domain;x_1,\ldots,x_n)}$ (we assume
the Green's function not to depend on the marked point $x$).
Suppose these are chosen so that for any conformal map
$\confmap: \domain \rightarrow \domain'$
sending $x,x_1,\ldots,x_n$ to $x',x_1',\ldots,x_n'$ we have
\begin{align*}
\label{eq: strict CI}
M_{\domain;x,x_1,\ldots,x_n}(z) \; = \; &
M_{\domain';x',x_1',\ldots,x_n'} (\confmap(z)) \qquad 
\textrm{ and } \\
C_{\domain;x_1,\ldots,x_n}(z_1,z_2) \; = \; &
C_{\domain';x_1',\ldots,x_n'} (\confmap(z_1), \confmap(z_2)) .
\tag{conf.inv.}
\end{align*}
In particular taking $\confmap = g_t$, the conditions
(\ref{M-eq}) and (\ref{C-eq}) require the processes
\begin{align}
\label{eq: M-ci}
&M_{\domain_0; X_t, g_t(x_1), \ldots, g_t(x_n)} (g_t(z))
\\
\label{eq: C-ci}
& M_{\domain_0; X_t, \ldots, g_t(x_n)} (g_t(z_1)) \,
M_{\domain_0; X_t, \ldots, g_t(x_n)} (g_t(z_2))
\; + \; C_{\domain_0;g_t(x_1),\ldots,g_t(x_n)}(g_t(z_1), g_t(z_2))
\end{align}
to be martingales. Since $(X_t)_{t \in [0,\stopping]}$
is a semimartingale and the flow $(g_t)_{t \in [0,\stopping]}$ is governed by
\eqnref{eq: Loewner},
computing the It\^o derivatives of the two processes is now easy.

Write first of all the It\^o diffusion of the driving process as
\begin{align}
\label{eq: general driving process}
\ud X_t = \ud W_{\kappa t} + \tau_{X_t} D_t \, \ud t
\end{align}
where $(W_t)_{t \geq 0}$ is a standard Brownian motion on $\bdry \domain_0$
and $\tau_x$ are positively oriented unit tangents to $\bdry \domain_0$ at $x$.
Then write $M_{\domain_0;x,x_1,\ldots,x_n}(z)$ as the imaginary part of an
analytic function $F(z;x,x_1,\ldots,x_n)$ on $\domain_0$, which we assume to
depend smoothly also on the marked points $x,x_1,\ldots,x_n \in \bdry \domain_0$.
The It\^o derivative of (\ref{eq: M-ci}) can be read from the
imaginary part of
\begin{align}
\label{eq: Ito derivative of F ci}
& \ud F(g_t(z);X_t,g_t(x_1),\ldots,x_n) \\
\nonumber
\; = \; &  \left( \sqrt{\kappa} \pder{x} F \right) \, \ud B_t
+ \left\{ V_{X_t} \big(g_t(z) \big) \pder{z} F
+ \frac{\kappa}{2} \pderS{x} F + D_t \pder{x} F
+ \sum_{j=1}^n V_{X_t} \big(g_t(x_j) \big) \frac{1}{\tau_{g_t(x_j)}} \pder{x_j} F \right\}
    \, \ud t ,
\end{align}
the right hand side being evaluated at 
$(g_t(z); X_t, g_t(x_1), \ldots, g_t(x_n))$.
\begin{remark} Note that in the above formula and in what follows $x$ and $x_i$
are points on the boundary and the derivatives $\pder{x}$ and $\pderS{x}$
should be understood as first and second derivatives with respect to length
parameter on the boundary (in the direction of the unit tangent $\tau$).
We do not assume analyticity with respect to those points.
Also, $B_t$ is a standard Brownian motion on $\bR$ such that
$\sqrt{\kappa} B_t$ is the length parameter of $W_{\kappa t}$.
\end{remark}

In view of \eqnref{eq: Ito derivative of F ci},
the condition of the mean (\ref{eq: M-ci}) being a local martingale is 
\begin{align*}
\label{eq: M-eq ci}
\im \left\{ V_{x}(z) \pder{z} F + \frac{\kappa}{2} \pderS{x} F + D_t \pder{x} F
+ \sum_{j=1}^n V_{x}(x_j) \frac{1}{\tau_{x_j}} \pder{x_j} F \right\} \; = \; 0 .
\tag{M-cond'}
\end{align*}
If this equation is satisfied then the drift of $M_t(z)$ vanishes and we have
\[ \ud M_t(z) = \sqrt{\kappa} \; \im \left(
\pder{x} F \big( g_t(z);X_t,\ldots,g_t(x_n) \big) \right) \, \ud B_t ,\]
and the It\^o derivative of (\ref{eq: C-ci}) significantly simplifies due to
the following
\begin{align*}
\ud \Big( M_t(z_1) \, M_t(z_2) \Big)
= \; & \big( \cdots \big) \, \ud B_t +
\kappa \; \imof{\pder{x} F \big(g_t(z_1); \ldots \big)} \;
\imof{\pder{x} F \big( g_t(z_2); \ldots \big)} \; \ud t .
\end{align*}
The condition (\ref{C-eq})
thus reduces to the following equation
\begin{align*}
\label{eq: C-eq ci}
& \der{t} \, C_{\domain_t; x_1, \ldots, x_n} (z_1, z_2)
\tag{C-cond'} \\
\; = \; & - \kappa \; \imof{\pder{x} F \big( g_t(z_1); X_t, \ldots, g_t(x_n) \big)} \;
\imof{\pder{x} F \big( g_t(z_2); X_t, \ldots, g_t(x_n) \big)} .
\end{align*}
In the strictly conformally covariant cases (\ref{eq: strict CI}), the verification
of the two basic conditions (\ref{M-eq}) and (\ref{C-eq}) therefore boils down
simply to Equations (\ref{eq: M-eq ci}) and (\ref{eq: C-eq ci}) as well as
appropriate boundedness of $M$.

\subsection{Example: chordal SLEs and Gaussian free fields}
\label{sec: SSExample}

\subsubsection*{Basic conditions for chordal SLE${}_4$ and GFF with jump-Dirichlet boundary conditions}
We will now illustrate the general idea in the example (\ref{eq: SS example})
of Schramm and Sheffield, by checking the conditions (\ref{M-eq}) and
(\ref{C-eq}) in this simplest case.

We take the upper half-plane as the starting domain $\Omega_0 = \bH$, and we
have two marked points $0$ and $\infty$. Our Loewner chain
(\ref{eq: Loewner H}) constructed by the vector fields $V_x(z)=\frac{2}{z-x}$
preserves one of them (infinity), and the other corresponds to the tip of the
curve. The driving process of chordal SLE${}_4$ is $X_t = \sqrt{\kappa} B_t$
with $\kappa=4$. Concretely, the equations (\ref{eq: M-eq ci})
and (\ref{eq: C-eq ci}) now have the following simple form:
\begin{align}
\label{MainEq1SS}
& \im \left( \frac{\kappa}{2} \pderS{x} F +  \frac{2}{z-x} \, \pder{z} F \right)
\; = \; 0 \qquad \textrm{ and } \\
\label{MainEq2SS}
& \der{t} C_{\Omega_t}(z_1,z_2) \; = \;
    - \kappa \; \im \Big(\pder{x} F \big( g_t(z_{1});X_t \big) \Big) \;
    \im \Big( \pder{x} F \big( g_t(z_{2});X_t \big) \Big) .
\end{align}

The harmonic function $M$ in $\bH$ determined by boundary conditions
(\ref{eq: SS example}) is $\frac{2\lambda}{\pi} \arg(z-x) - \lambda$, hence
$F(z;x) = \frac{2\lambda}{\pi} \log(z-x) - \lambda$ and an easy calculation
confirms the validity of Equation (\ref{MainEq1SS}) when $\kappa = 4$.
The Dirichlet Green's function in the half-plane is explicitly
\begin{equation}
\label{DirGreen}
C(z_1,z_2) \; = \; -\frac{1}{2 \pi} \; \re \log \Big( \frac{z_1-z_2}{z_1-\cc{z}_2} \Big).
\end{equation}
Applying conformal invariance of $C$, (\ref{eq: C-ci}), and
computing the time derivative of $C(g_t(z_1),g_t(z_2))$, we find that
(\ref{MainEq2SS}) also holds true provided that the jump size is adjusted
to the value found by Schramm and Sheffield, $\lambda = \pm \sqrt{\frac{\pi}{8}}$.

\begin{remark}
\label{rem: rigid}
The above calculation is rather rigid in the following sense.
Suppose we want to find some coupling of SLE with a free field whose
covariance is given by the Dirichlet Green's function (\ref{DirGreen}).
Then the equation (\ref{MainEq2SS}) in fact determines the function
$F$ up to a sign and an additive constant. One then only has to check that
such $F$ is a martingale for the SLE,
i.e. that Equation (\ref{MainEq1SS}) is satisfied.
\end{remark}

\begin{remark}
The right hand side of (\ref{MainEq2SS}) is
$- \frac{16 \lambda^2}{\pi^2} \; \im \big( \frac{1}{z_{1}-x} \big) \;
\im \big( \frac{1}{z_{2}-x} \big)$.
Terms in this product have invariant meaning. Namely, they are multiples
of Poisson's kernel with zero Dirichlet boundary conditions. This is
a general phenomenon and a consequence of a formula of Hadamard type which
we discuss in \secref{sec: Hadamard}.
\end{remark}

\subsubsection*{Modification to chordal SLE${}_\kappa$ for $\kappa \neq 4$}
There is a way to save the
validity of the basic conditions for chordal SLE${}_\kappa$, $\kappa \neq 4$,
if one relaxes the assumption (\ref{eq: strict CI}) of strict
conformal invariance of $M$. By Remark \ref{rem: rigid} the choice of the
Dirichlet Green's function together with Equation (\ref{MainEq2SS})
implies we should take the same $F$ as before but with
$\lambda = \lambda_\kappa = \pm \sqrt{\frac{\pi}{2\kappa}}$.
However, Equation (\ref{MainEq1SS}) fails for general $\kappa$
and we have instead
\[ \im \left( \frac{\kappa}{2} \pderS{x} F
        + \frac{2}{z-x} \, \pder{z} F \right)
\; = \; \frac{(4-\kappa) \lambda_\kappa}{\pi} \;
    \im \Big( \frac{1}{(z-x)^2} \big) \; \neq \; 0 .
\]
We therefore adjust the definition of the mean $M_{\domain_t}(z)$ for
the field in the new domain $\Omega_t$ as follows
\begin{align}
\label{eq: ad hoc term}
M_{\domain_t} (z) \; = \; \im \Big( F \big( g_t(z); X_t \big) + E_t(z) \Big) ,
\end{align}
where the extra term $E_t(z)$ is taken to be the integral of the missing part
\begin{align}
\label{eq: ad hoc term 2}
\der{t} E_t(z) \; = \; \frac{(\kappa-4) \lambda_\kappa}{\pi} \; \frac{1}{(g_t(z)-X_t)^2} ,
\qquad E_0(z) = 0 .
\end{align}
This guarantees that $M_{\domain_t}(z)$ are local martingales. Condition
(\ref{M-eq}) follows for appropriate stopping times $\stopping$,
and since the added term $E_t(z)$ is of finite variation, the computation
leading to Equation (\ref{MainEq2SS}) remains unchanged and implies (\ref{C-eq}).

The definition (\ref{eq: ad hoc term 2}) can be explicitly integrated to give
\begin{align} \label{eq: ad hoc term 3}
E_t(z) \; = \; \frac{(4-\kappa) \lambda_\kappa}{2\pi} \; \log g_t'(z) ,
\end{align}
simply using $\der{t} g_t'(z) = \frac{-2 g_t'(z)}{(g_t(z)-X_t)^2}$.
In particular, $\im (E_t(z))$ is determined by the domain $\domain_t$ only
and could be interpreted as a multiple
of the harmonic interpolation of the argument of the tangent vector $\tau$ 
of $\bdry \domain_t$ (``winding of the boundary'') if the boundary
would be smooth. In Appendix \ref{sec: commutation} we show that for general
boundary conditions the mean $M_{\domain_t}$ defined as in
(\ref{eq: ad hoc term}) can depend on the full history of the Loewner chain
$(g_s)_{0 \leq s \leq t}$ and not be determined by the domain $\domain_t$ only.

We remark also that the additional term
(\ref{eq: ad hoc term 3}) is what the Coulomb gas formalism of conformal
field theory dictates in the presence of a background charge which 
modifies the central charge $c$ to its correct value
$c(\kappa) = 1 - 6 \, (\frac{\kappa-4}{2 \sqrt{\kappa}})^2$.

\subsection{Basic equations imply coupling}
\label{sec: coupling}

\subsubsection*{Definition of the free fields}
Let us now give a precise definition of our free fields $\Phi$. It is common to define
them as random tempered distributions, although they are almost surely somewhat more regular
objects. 
We denote by $\rapdec$ the Schwarz class of functions of rapid decrease on
$\bC = \bR^2$ and by $\tempdist$ the tempered distributions.
Define the function $W : \rapdec \rightarrow \bC$ which will be the characteristic
function of $\Phi$
$$ W(f) \; = \; \exp \left( \ii \int_\domain M(z) f(z) \ud z -
    \frac{1}{2} \iint_{\domain \times \domain} f(z) C(z,w) f(w) \, \ud z \, \ud w\right) .$$
We clearly have $W(0)=1$.
All of our choices of functions $M$ and $C$ will satisfy the properties
\begin{itemize}
\item 
The function $M : \domain \rightarrow \bR$ is locally integrable
and has at most polynomial growth at infinity
\item 
The function $C$ is locally integrable
and has at most polynomial growth at infinity
\end{itemize}
which imply that $W$ is continuous. Furthermore, 
all of our choices of $C$ will have the property
\begin{itemize}
\item 
For all $f_1, \ldots, f_n \in \rapdec$ the $n \times n$ real matrix with entries
$C_{j,k} = \iint f_j(z) C(z,w) f_k(w) \, \ud z \, \ud w$ is positive semi-definite
\end{itemize}
so a standard argument shows that for all
$\zeta_1, \ldots, \zeta_n \in \bC$ and $f_1, \ldots, f_n \in \rapdec$ we have
$\sum_{j,k} \zeta_j \cc{\zeta}_k W(f_j-f_k) \geq 0$. These conditions guarantee,
by Minlos' theorem, that $W$ is indeed a characteristic function of a
probability measure 
on $\tempdist$,
which is our definition of the law of the massless free field with mean $M$ and
covariance $C$.

It is evident from the definition of $W$ that
the free field is almost surely supported on $\cl{\domain}$.

\subsubsection*{Coupling}

\begin{theorem} \label{thm: coupling}
Let $A \subset \domain$ be compact and $B \subset \Omega$ open neighborhood
of $A$. Let $(g_t)_{t \geq 0}$ be a random Loewner chain of hulls
$(\hull_t)_{t \geq 0}$ 
and suppose that $\stopping$ is a stopping time
for which $\cl{\hull}_\stopping \cap (B \cup \set{x_1, \ldots, x_n}) = \emptyset$.
Denote the tip of the hull at time $t$ by $\tilde{x}(t)$ and the complement by
$\domain_t = \domain \setminus \hull_t$.
Assume conditions (M-mgale) and (C-mgale).
Then the random Loewner chain $(g_t)_{t \in [0,\stopping]}$
can be coupled with a free field $\wtil{\Phi}$
defined on $A$ such that the following holds.
\begin{itemize}
\item Let $\stoppingBis \leq \stopping$ be a stopping time. Conditionally on
$(g_s)_{0 \leq s \leq \stoppingBis}$,
the law of $\wtil{\Phi}$ is the restriction to $A$
of the free field corresponding to the domain $\domain_\stoppingBis$, that is
the free field with mean $M_{(\domain_\stoppingBis,\tilde{x}(\stoppingBis),x_1,\ldots,x_n)}$
and covariance $C_{(\domain_\stoppingBis,\tilde{x}(\stoppingBis),x_1,\ldots,x_n)}$.
\end{itemize}
\end{theorem}
\begin{proof}
The theorem will be proved by showing that for any test function $f$
the expectation
$\EX_{\mathrm{GFF}} [\exp( \ii \,\bra \wtil{\Phi}_t, f\ket )]$
is a martingale, where $\wtil{\Phi}_t$ has the law of the free
field in $\domain_t = \domain \setminus \hull_t$.

Denote by $M_t$ the mean associated to the domain
$\domain_t$
with marked points $\tilde{x}(t), x_1, \ldots, x_n$, and by $C_t$ the
covariance associated to that domain.
Define $\wtil{\Phi}$ by sampling a free field in $\domain_\stopping$
with mean and covariance $M_\stopping$ and $C_\stopping$, and then restricting to $A$.

Given $f \in \rapdec$, $\supp{f} \subset A$, we define first of all the process
$$ L_t \; = \; \int_A M_t(z) f(z) \ud z .$$
By the assumption (M-mgale)
$(L_t)_{t \in [0,\stopping]}$ is a bounded continuous martingale.
Its quadratic variation follows from assumption (C-mgale)
$$ \bra L, L \ket_t \; = \;
\iint_{A \times A} f(z) \Big( C_0(z,w) - C_t(z,w) \Big) f(w)
\, \ud z \, \ud w .$$

For any $f \in \rapdec$ such that $\supp{f} \subset A$, define the random process
$(\wtil{W}_t(f))_{t \in [0,\stopping]}$ by 
$$ \wtil{W}_t(f) \; = \; \exp \left( \ii \int_A M_t(z) f(z) \ud z
    - \frac{1}{2} \iint_{A \times A}  f(z) C_t(z,w) f(w) \, \ud z \, \ud w \right) . $$
Note that $\wtil{W}_t(f)$ is up to a multiplicative constant the exponential
martingale $\exp \Big( \ii \, L_t + \frac{1}{2} \bra L,L \ket_t \Big)$, so
in particular it is a bounded martingale.

It is now easy to describe the law of the random distribution $\wtil{\Phi}$
conditionally on $(g_s)_{0 \leq s \leq t}$. The law is encoded in
the characteristic function
$\EX [ \exp( \ii \,\bra \wtil{\Phi}, f \ket ) \, | \, \sF_t ] $ .
At time $t = \stopping$ this is exactly the characteristic function of
$\wtil{\Phi}$, that is $\wtil{W}_\stopping(f)$. By construction it is a
bounded martingale and therefore coincides with
$$\EX [ \exp( \ii \,\bra \wtil{\Phi}, f \ket ) \, | \, \sF_t ] \; = \;
\wtil{W}_t(f) . $$
Since $\wtil{W}_t$ is by construction the characteristic function of
the free field with mean and covariance $M_t$ and $C_t$ the assertion follows.
\end{proof}

\subsection{Hadamard's variational formulas for Loewner chains}
\label{sec: Hadamard}
Hadamard's 
formula gives the variation of Green's function in a smooth domain when the
boundary changes in a smooth way.  In this section we prove a version of
Hadamard's formula for Loewner chains. The need for this stems from
the second basic condition for coupling --- in \mbox{\eqnref{eq: C-eq ci}}
we need the derivative of Green's functions in domains $\domain_t$
with respect to the time $t$ of the Loewner chain.
\begin{theorem}
\label{theorem: Hadamard}
Let $(g_t)_{t \geq 0}$ be a Loewner chain in a
simply or doubly connected domain as in \secref{sec: Loewner chains},
and let $\domain_t = \domain_0 \setminus \hull_t$.
Let $G_{\domain_t}(z_1,z_2)$ be the zero Dirichlet
boundary valued Green function in $\domain_t$. Then   
\begin{equation}
\label{eq: Hadamard thm}
\der{t} G_{\domain_t}(z_1,z_2) \big|_{t=0} \; = \;
- 2 \pi \; P_{\domain_0}(X_0,z_1) \; P_{\domain_0}(X_0,z_2) ,
\end{equation}
where $P_{\domain}$ is the Poisson kernel in $\domain$.
\end{theorem}
\begin{proof}
Fix the point $z_2$. The difference
$\Gamma_{z_2}(z_1):=G_{\domain_0}(z_1,z_2)-G_{\domain_t}(z_1,z_2)$
is harmonic in $\domain_t$ as function of $z_1$ and thus can be
represented as the integral of its boundary values against the
harmonic measure:
\begin{equation}
\Gamma_{z_2}(z_1) \; = \; \int_{\partial \domain_t} G_{\domain_0}(z,z_2) \;
    \ud \harmmeas^{\domain_t}_{z_1}(z)
\; = \; \int_{\bdry \hull_t} G_{\domain_0}(z,z_2) \; \ud \harmmeas^{\domain_t}_{z_1}(z),
\label{Had_difference}
\end{equation}
since the boundary values are zero everywhere but on $\partial \hull_t$.

By conformal invariance we may assume that $X_0=0$,
$\domain_0 \subset \bH$ and $\domain_0$ coincides with the upper half-plane $\bH$
in some neighborhood of $X_0=0$. If $x+iy = z \in \partial \hull_t$, then
\[ G_{\domain_0}(z,z_2) \; = \; G_{\domain_0}(x,z_2) + y \; \nder G_{\domain_0}(x,z_2)+ o(y), \;y\rightarrow 0.  \]
The first term in the right-hand side is equal to $0$,
and the normal derivative of the Green's function is the Poisson kernel
$P_{\domain_0}(x;z_2)$, which is roughly the same as $P_{\domain_0}(0;z_2)$.
More precisely, one has
\[ G_{\domain_0}(z,z_2) \; = \; y \, P_{\domain_0}(0;z_2)+ y \, O(x)+o(y). \]
Hence, (\ref{Had_difference}) reads
\[ \Gamma_{z_2}(z_1) \; \approx \;
P_{\domain_0}(0;z_2) \; \int_{\partial \hull_t} \im (z) \; \ud \harmmeas^{\domain_t}_{z_1}(z)
\; =: \; P_{\domain_0}(0,z_2) \; \Psi(z_1). \]
The notation "$\approx$"  means that the ratio of two expressions tends to $1$
as the size of the hull tends to zero. Now, take a small $r>0$ such that
$\hull_t$ is inside the demi-circle $T_r(0)$ of radius $r$ around $0$.
Denote $\domain^{(r)} := \domain_0\setminus B_r(0)$. Write $\Psi(z_1)$ as 
\begin{equation}
\label{HadPsi} 
\Psi(z_1)=\int_{T_r(0)} \Psi(z) \; \ud \harmmeas^{\domain^{(r)}}_{z_1}(z).
\end{equation}
We are going to factor out a term that captures the dependence of the
latter integral on $z_1$. To this end, we apply the map
$\psi_r(z):=z+\frac{r^2}{z}$ which maps $\mathbb{H} \setminus B_r(0)$
onto $\mathbb{H}$ (and $\domain_0$ onto some domain $\psi(\domain_0)$).
Now, conformal invariance of the harmonic measure yelds 
\[ \ud \harmmeas^{\domain^{(r)}}_{z_1}(z) \; = \;
\ud \harmmeas^{\psi_r(\domain^{(r)})}_{\psi_r(z_1)}(\psi_r(z)) \; = \;
P_{\psi_r(\domain)}(\psi_r(z);\psi_r(z_1)) \; \ud x . \] 
Since $\psi_r(z)-z$ is small when $r$ is small, we have 
\begin{equation}
\label{PoissonClose}
P_{\psi_r(\domain)}(\psi_r(z);\psi_r(z_1)) \; \approx \;
P_{\psi_r(\domain)}(0;\psi_r(z_1)) \; \approx \; P_{\domain}(0;z_1) .
\end{equation}
Hence the equation (\ref{HadPsi}) reads
\[ \Psi(z_1) \; \approx \; P_{\domain}(0,z_1) \;
    \int^{\pi}_{\theta=0} \Psi(r e^{\ii \theta}) 2 \sin (\theta) r \; \ud \theta. \]
The integral on the right-hand side is by definition equal to
$\pi \, L^\domain_{K_t,r}$, where $L^\domain_{K_t,r}$ is the 
local half-plane capacity, see (\ref{lhcap}), and we apply
\propref{prop: hloc} to finish the proof.
\end{proof}

It is easy to generalize this theorem to other boundary conditions.
One possible generalization is as follows. Let the boundary of the
domain $\domain=\domain_0$ consist of several connected components,
that in turn are divided to several arcs each. Without loss of
generality, assume $\partial \domain$ to be piecewize smooth,
and let $\tilde{G}_{\domain}(z_1,z_2))$ be the Green's function with
zero Dirichlet boundary conditions on some of those arcs and Neumann
boundary conditions on others. Let $(\domain_t)$ be a family of
domains defined by a Loewner chain (the setup for the chain being
analogous to that of section \ref{sec: Loewner chains}, with residue
of absolute value $2$ at the marked point). We demand that the point
of growth $X_0 \in \partial \domain$ of the Loewner chain would
belong to the ``Dirichlet'' part of the boundary, and by definition
$\tilde{G}_{\domain_t}(z_1,z_2))$ assumes zero Dirichlet boundary
values on $\hull_t$. Then we have the following proposition:
\begin{proposition}
\label{prop: HadGeneral}
\begin{equation}
\label{Hadamard}
\der{t} \tilde{G}_{\domain_t}(z_1,z_2)) \big|_{t=0} \; = \;
-2\pi \; \tilde{P}_\domain(X_0; z_1) \, \tilde{P}_\domain(X_0; z_2).
\end{equation}
where $\tilde{P}_\domain$ is the Poisson kernel with the same
boundary conditions as $\tilde{G}$.
\end{proposition}
\begin{proof}
The proof literally repeats the one of \thmref{theorem: Hadamard};
there are only two places where we have
used a specific nature of the boundary conditions far away from
the point $X_0$. One is the continuity or the Poisson kernel with
respect to small variations of the domain
(equation (\ref{PoissonClose})). This is also clear in the
present case. Another one is the definition of $\lochcap(\hull_t)$,
namely, the boundary conditions for $\Psi$ in (\ref{lhcap}). It
is clear, however, that they can be replaced by Neumann ones far
from the point $X_0$, and the difference at the distance $r$
from $a$ is of order $o(r^2)$, which is negligible when
computing $\partial_t \lochcap(\hull_t) |_{t=0}$.
\end{proof}

\section{Various boundary conditions in simply connected domains}
\label{sec: simply connected}

\subsection{SLE${}_4$ in the strip and Riemann-Hilbert boundary conditions}
\label{sec: strip RH}
In this subsection, we develop a coupling of SLE${}_4$ and GFF in the
following situation. We take $\domain$ to be a simply-connected domain with three marked
points on the boundary $x_0,x_1,x_2$ dividing the boundary into three arcs
$l_{12}, l_{01}, l_{20}$. The mean $M (z) = M_{\domain;x_0,x_1,x_2}(z)$
of the field will be a harmonic function determined by the boundary conditions
\begin{align} \label{eq: Riemann-Hilbert M}
\left\{ \begin{array}{ll}
M(z) = - \lambda & \textrm{ for $z \in l_{01}$} \\
M(z) = \lambda & \textrm{ for $z \in l_{20}$} \\
\alpha \, \nder M(z) + \beta \, \tder M(z) = 0 & \textrm{ for $z \in l_{12}$.}
\end{array} \right.
\end{align}
The third condition can be reformulated in the following way:
if $M(z) = \im F(z)$, then on $l_{12}$ the derivative of $F$ in the
direction of the boundary has a constant argument modulo $\pi$.
If at some point of the arc $l_{12}$ the function $F$ vanishes,
this implies that $F$ itself has the same argument (modulo $\pi$) on
$l_{12}$.
As the covariance $C(z_1,z_2) = C_{\domain;x_1,x_2}(z_1,z_2)$ we take
the Green's function in $\domain$ having zero Dirichlet boundary
conditions on $l_{20}$ and $l_{01}$
and the above type of Riemann-Hilbert boundary conditions on $l_{12}$:
for all $z_2 \in \domain$ we require
\begin{align} \label{eq: RH for C}
\left\{ \begin{array}{ll}
C(\cdot,z_2) = 0 \quad & \textrm{ on  $l_{20} \cup l_{01}$} \\
\alpha \, \nder C(\cdot,z_2) + \beta \, \tder C(\cdot,z_2) \equiv 0
\quad & \textrm{ on  $l_{12}$.} \end{array} \right.
\end{align}
These boundary conditions are conformally invariant in the sense of
\eqnref{eq: strict CI}, so the essential part of establishing a coupling
consists of verifying \eqnDref{eq: M-eq ci}{eq: C-eq ci}.
We already remark that Dirichlet and Neumann boundary conditions on $l_{12}$
are particular cases corresponding to vanishing $\alpha$ and vanishing $\beta$,
respectively,
and after having made the coupling explicit we return to comment on an
interpolation between the two.

The convenient choice of Loewner chain for the domains $(\domain;x_0,x_1,x_2)$
with three marked boundary points is to keep $x_1$ and $x_2$ as fixed points.
We therefore take
the initial domain to be the strip, $\domain_0 = \bS$, with $x_1$ and $x_2$
at $+\infty$ and $-\infty$ respectively, and we use (\ref{eq: Loewner S})
to encode the growth process. We furthermore choose $X_0=x_0=0$.
Below $F(\cdot;x)$ denotes an analytic function in $\bS$
whose imaginary part is the harmonic function $M_{\bS;x,+\infty,-\infty}$
determined by (\ref{eq: Riemann-Hilbert M}).

As the marked points $x_1$, $x_2$ are chosen to be fixed by the Loewner
flow, the basic equations (\ref{eq: M-eq ci}) and (\ref{eq: C-eq ci}) have
a simple form
\begin{align}
\label{MainEq1Strip}
& \im \left\{ 2 \; \pderS{x} F(z;x) + \coth \big( \frac{z-x}{2} \big) \; \pder{z} F(z;x)
    + D_t \; \pder{x}  F(z;x) \right\} \; = \; 0 \qquad \textrm{ and } \\
\label{MainEq2Strip}
& \der{t} C_{\domain_t}(z_1,z_2) = - 4 \; \im \left( \pder{x} F(g_t(z_{1});X_t) \right)
\; \im \left( \pder{x} F(g_t(z_{2});X_t) \right) .
\end{align}
\propref{prop: HadGeneral} combined with conformal invariance readily
gives the expression
\[ \der{t} C_{\domain_t}(z_1,z_2) \; = \;
- 2 \pi \; \tilde{P}(X_t ; g_t(z_1)) \; \tilde{P}(X_t ; g_t(z_2)), \]
where $\tilde{P}$ is the Poisson kernel in $\mathbb{S}$ having the same
 boundary conditions as the Green's function (\ref{eq: RH for C}).
As before, \eqnref{MainEq2Strip} therefore determines
$\pder{x} F(z;x)$ up to a sign and a constant
\begin{align} \label{eq: F generic formula}
\pder{x} F(z,x) \; = \;
\pm \ii \; \sqrt{\frac{\pi}{2}} \; \tilde{S}_x(z) + \textrm{real constant},
\end{align}
where $\tilde{S}_x(z)$ is the Schwarz kernel corresponding to the present
boundary conditions --- an analytic function in $\bS$ such that
$\re ( \tilde{S}_x(z) ) = \tilde{P}_x(z)$.

We should then verify (\ref{MainEq1Strip}). Note first that our function
$F$ is invariant under shifts
\[ \pder{x} F + \pder{z} F = 0, \]
and hence (\ref{MainEq1Strip}) reads equivalently
\begin{equation}
\label{StripHarmonic}
\im \left\{ 2 \; \pderS{x} F(z;x) - \coth(\frac{z-x}{2}) \; \pder{x} F(z;x)
    + D_t \; \pder{x} F(z;x) \right\} \; = \; 0 .
\end{equation}
This identity could be checked for correctly chosen $D_t$
by a direct calculation using an explicit expression for $\tilde{S}$,
but we prefer an argument which identifies the drift $D_t$ in a way that
generalizes directly to other cases where explicit expressions may in
practise be unavailable. A similar technique was used by Zhan
in the context of loop-erased random walks in multiply
connected domains \cite{Zhan-thesis}

The function on the left-hand side of (\ref{StripHarmonic}) is harmonic
in $\mathbb{S}$, it is zero on $\bR$ and bounded apart from a possible
singularity at $x$. On the upper part of the boundary, the first and
third terms clearly satisfy the $(\alpha, \beta)$ Riemann-Hilbert
boundary condition. In order to prove the same condition for the second
term, recall that $\pder{x} F$ was defined up to a real constant.
If we now choose that constant so that $\re (\pder{x} F) = 0$ at $-\infty$,
then clearly $\pder{x} F = 0$ at $-\infty$, and the Riemann-Hilbert
boundary condition for $\im (\pder{x} F)$ can be stated in the form
that $\arg \pder{x} F$ modulo $\pi$ is fixed on $\bR + \ii \pi$.
Since $\coth(\frac{z-x}{2})$ is purely real, multiplication by it doesn't
harm this condition. 

It remains to prove that singularities of the left-hand side of
(\ref{StripHarmonic}) at the point $x$ actually cancel out. Expansions
at $x$ for the Schwarz kernel and the Loewner vector field give
\begin{align*}
\partial_x F(z;x) \; = \; & \frac{C}{z-x} + C \, \mu + o(1) , \\
\coth(\frac{z-x}{2}) \; = \; & \frac{2}{z-x}+o(1),
\end{align*}
where $C$ and $\mu$ are real since the Schwarz kernel $\tilde{S}_x(z)$
is purely imaginary on the real line.
Hence, the left-hand side of (\ref{StripHarmonic}) is bounded
if and only if
\[ D_t \; \equiv \; 2 \, \mu , \]
which determines the drift $D_t$ of the driving process
(\ref{eq: general driving process}) and establishes the condition
(\ref{eq: M-eq ci}) for the correctly chosen drift.

In order to find $\mu$ in terms of $\alpha$ and $\beta$, we need
the explicit formula for the function $\pder{x} F$. Note that 
for $-\frac{1}{2} < \theta < \frac{1}{2}$ the expression
\begin{align} \label{eq: explicit RH Schwarz}
\tilde{S}_x(z) \; = \;
\frac{\ii}{2 \pi}\frac{e^{\theta (z-x)}}{\sinh(\frac{z-x}{2})}
\end{align}
gives a Schwarz kernel in $\mathbb{S}$ satisfying 
\[ \arg \partial_\tau \tilde{S} = \pi \theta \; \mod \pi \qquad
\textrm{ on $\bR + \ii \pi .$} \]
so we find that for such boundary conditions $\mu=\theta$.
We have proven the following proposition.
\begin{proposition}
Choose $\lambda = \sqrt{\pi/8}$ and
\[ \alpha = \cos(\pi \theta) , \qquad \beta = -\sin(\pi \theta) \]
and let $\Phi$ be the Gaussian free fields 
with means $M_{\domain; x_0,x_1,x_2}(z)$ determined by boundary
conditions (\ref{eq: Riemann-Hilbert M}), and covariances
$C_{\domain; x_1, x_2}$ determined by (\ref{eq: RH for C}).
Then $\Phi$ are coupled, in the sense of \thmref{thm: coupling},
with the SLE${}_4(\rho)$ in $\bS$ with $\rho = 2 \theta - 1$.
\end{proposition}
\begin{remark}
Free fields and SLEs are conformally invariant if we allow for
(random) time reparametrizations of the Loewner chains, so the given coupling
works in any other domain $(\domain; x_0,x_1,x_2)$, too.
\end{remark}
\begin{remark}
Both cases $\theta \rightarrow \pm \half$ correspond to Dirichlet boundary
conditions also on $l_{12} = \bR + \ii \pi$. Correspondingly, the curves become
just chordal SLE${}_4$ in the strip from $0$ to $\pm \infty$, and
these cases can be seen as mere coordinate changes of the case of
Schramm \& Sheffield discussed in \secref{sec: SSExample}.
\end{remark}
\begin{remark}
The symmetric value $\theta = 0$ corresponds to Neumann boundary conditions
on $\bR + \ii \pi$.
The drift $D_t$ then vanishes and the curve is a dipolar SLE${}_4$.
It appears that this case was first conjectured in \cite{BBH-dipolar}.
\end{remark}
\begin{remark}
As $\theta$ varies from $-\half$ to $\half$, the free fields and the curves
interpolate between the above cases. This was suggested in
\cite{Kytola-SLE_kappa_rho}, where 
$\tilde{S}_x(z)$ was also used to give a formula for left passage probability
of the SLE${}_4(\rho)$ curve.
\end{remark}

One would like, as in the chordal case, to extend the coupling to
$\kappa \neq 4$. Again, \eqnref{MainEq2Strip} and Hadamard's
formula for $C$ leave us essentially no choice but
$\pder{x} F(z;x) = 2 \ii \, \lambda_\kappa \, \tilde{S}_x(z)$ with
$\lambda_\kappa = \sqrt{\frac{\pi}{2 \kappa}}$.
\eqnref{MainEq1Strip} then fails, giving instead
\[ \im \left\{ \frac{\kappa}{2} \; \pderS{x} F(z;x)
    + \coth \big( \frac{z-x}{2} \big) \; \pder{z} F(z;x)
    + D_t \; \pder{x}  F(z;x) \right\} \; = \;
(\kappa-4) \lambda_\kappa \; \im \left( \ii \, \pder{x} \tilde{S}_x(z) \right)
\; \neq \; 0 \]
As in (\ref{eq: ad hoc term}), we could try to save the basic conditions
by adding a non-conformally invariant term $E_t$ to the mean of the field:
$M_{\domain_t}(z) = \im \big( F(g_t(z);X_t) + E_t(z) \big)$, now
taken to be
\begin{align} \label{eq: ad hoc strip}
E_t(z) \; = \; (4-\kappa) \lambda_\kappa \;
    \int_0^t \left( \ii \, \pder{x} \tilde{S}_{X_s}(g_s(z)) \right) \, \ud s .
\end{align}
One observes that $E_t$, thus defined, satisfies the following properties: 
\begin{itemize}
\item $\im (E_t)$ has the same Riemann-Hilbert boundary conditions as
$\im (F)$ on $\bR + \ii \pi$
\item $\im (E_t) \equiv 0$ on $\bdry \domain_t \cap \bR$
\item If $z \in \bdry \hull_t$ for some $t$, then $\im (E_s(z)) = \im (E_t(z))$
for all $s>t$ unless the point $z$ is swallowed by time $s$.
Thus, the boundary value of $\im E$ on the curve is determined at the
instant the point becomes a part of the boundary. Note that this property
also held for the winding boundary conditions (\ref{eq: ad hoc term 3})
which generalized the chordal coupling to $\kappa \neq 4$.
\end{itemize}
Despite the above properties, there is a crucial difference to the case
of jump-Dirichlet boundary conditions: the mean (\ref{eq: ad hoc term})
will be determined by the domain only if the commutation condition of
Appendix \ref{sec: commutation} is satisfied --- and
for \eqnref{eq: ad hoc strip} it is not.

\subsection{More marked points}
In this section, we show how to compute the driving process of the SLE${}_4$
variant coupled with free field whose boundary conditions change also
at additional marked points
$-\infty+ \ii \, \pi = x_0, x_1, x_2, \dots,x_{n+1} = \infty+\ii \, \pi$
on the upper boundary of $\mathbb{S}$. In our example, the mean of the
field will satisfy the following boundary conditions:
\begin{align}
\left\{ \begin{array}{ll}
M(z,x,x_1,\dots,x_n)=-\lambda & \textrm{ for } z \in (x,+\infty) \\
M(z,x,x_1,\dots,x_n)=+\lambda & \textrm{ for } z \in (-\infty,x) \\
M(z,x,x_1,\dots,x_n)  \textrm{ obeys BC${}_i$ } & \textrm{ for }
    z \in l_i := (x_i, x_{i+1}) \subset \bR + \ii \pi
\end{array} \right.
\end{align}
\begin{itemize}
\item Here BC${}_i$ may stand either for constant Dirichlet condition
$M \equiv \lambda_i$, or zero Neumann boundary condition $\nder M \equiv 0$.
\end{itemize}
The covariance $C(z_1,z_2;x,x_1,\dots,x_n)$ is taken to have zero Dirichlet
boundary conditions on $\bR$, and BC'${}_i$ on $l_i$, where BC'${}_i$
stands for the homogeneous condition corresponding to BC${}_i$.
We've only given the mean and covariance in 
$(\bS; x, -\infty, x_1, \ldots, x_n, +\infty)$, but it is
understood that the definitions are transported to other domains with
marked points by \eqnref{eq: strict CI}.

The initial position of the growth is $X_0=0$. Let $\tilde{M}$
be the harmonic conjugate to $M$ normalized to be equal to $0$ at
$-\infty$, and let $\tilde{S}_x(z)$ be the Schwarz kernel with
BC'${}_i$ boundary conditions on the corresponding segments of
the upper boundary and with the same normalization at $-\infty$.

We have the following proposition:
\begin{proposition}
For $\lambda=\sqrt{\frac{\pi}{8}}$, there exist a unique function
$D(x,x_1,x_2,\dots,x_n)$ such that the SLE${}_4$ variant defined by
(\ref{eq: Loewner S}) with the driving process 
\[ \ud X_t \; = \; 2 \; \ud B_t + D(X_t, g_t(x_1), \dots, g_t(x_n)) \; \ud t \]
is coupled with the GFF described above.
The function $D(x,x_1,x_2,\dots,x_n)$ is given by
\begin{equation}
\label{defineD}
D(x,x_1,\dots,x_n) \; = \; 2 \; \mu(x, x_1, \dots, x_n)
    - 2 \; \sum_{i=0}^{n} \pder{x_i} \tilde{M}(x,x,x_1,\dots,x_n) ,
\end{equation}
where $\mu$ is the second coefficient in the expansion
at $z=x$ of the Schwarz kernel $\tilde{S}_x(z)$
\[ \mu := \frac{\pi}{\ii} \lim_{z\rightarrow x}
    \left( \tilde{S}_x(z) - \frac{\ii}{\pi(z-x)} \right) . \]
\end{proposition}
\begin{proof}
Hadamard's formula implies that \eqnref{eq: C-eq ci}
will hold provided that when we write $M = \im (F)$ the function
$F$ satisfies $\pder{x} F \, = \, 2 \ii \lambda \, \tilde{S}$, where
$\tilde{S}$ is the Schwarz kernel with corresponding boundary
conditions. The first equation (\ref{eq: M-eq ci}) now reads
\begin{equation}
\label{MainEq1StripMP}
\im \left\{ 2\, \pderS{x} F + \coth(\frac{z-x}{2}) \; \pder{z} F
    + \sum_i \coth(\frac{x_i-x}{2}) \; \pder{x_i} F
    + D_t \; \pder{x} F \right\} \; = \; 0 .
\end{equation}

Obviously the function in the parentheses in (\ref{MainEq1StripMP})
is purely real when $z \in \bR \setminus \{x\}$. We show
that it satisfies the homogeneous BC'${}_i$ boundary conditions on
the upper part of the boundary, and that for appropriate choice of
$D_t$ the singularities at $x$ cancel out. It is clear that if the
function $\im (F)$ satisfies Dirichlet or zero Neumann boundary
conditions on $l_i=(x_i,x_{i+1})\subset \bR + \ii \pi$,
then $\im (\pderS{x} F)$, $\im (\pder{x_i} F)$ and $\im (\pder{x} F)$
satisfy corresponding homogeneous conditions. The function $\pder{z} F$
is purely real where BC${}_i$ is Dirichlet and purely imaginary where
BC${}_i$ is Neumann. Multiplication by real constant does
not affect homogeneous Dirichlet or Neumann boundary conditions,
and multiplication by the real function $\coth(\frac{z-x}{2})$
does not change the argument of $\pder{z} F$ modulo $\pi$. So all terms
in (\ref{MainEq1StripMP}) satisfy BC'${}_i$ on $l_i$.

Our function $F$ is invariant under simultanious translation of all arguments,
i.e. \[ \pder{z} F \;= \; - \pder{x} F - \sum_i \pder{x_i}F . \]
So, we rewrite the equation (\ref{MainEq1StripMP}) as 
\begin{equation}
\label{StripFunnyEq}
\im \left\{ 2 \; \pderS{x} F - \coth(\frac{z-x}{2}) \; \pder{x} F
    - \coth(\frac{z-x}{2}) \; \sum_i \pder{x_i} F
    + \sum_i \coth(\frac{x_i-x}{2}) \, \pder{x_i} F +
    D_t \; \pder{x} F \right\} \; = \; 0
\end{equation}
Note that $\partial_{x_i} F$ might have a singularity at $x_i$ ---
however, it can only be of order $O((z-x_i)^{-1})$, so in the above
expression these singularities cancel out, and the function is
bounded near $x_i$'s. It remains to handle the singularity at $x$.
To do so, note that the expansion of $\pder{x} F$ at $x$ is
\[ \pder{x} F \; = \; \frac{C}{z-x} + C \, \mu + o(1) , \]
where $C$ is a real constant and $\mu$ is as specified in the
statement. Hence the second-order singularities, which only come from the
first two terms, cancel out. The first-order singularities come
from the second, the third and the last term in (\ref{StripFunnyEq}).
Clearly, there is a unique choice of $D_t$, specified in the
statement of the proposition, for which they also cancel out.
\end{proof}
\begin{remark}
If one wishes, one may allow some of BC${}_i$'s be Riemann-Hilbert
boundary conditions. The proof is similar to the above one, and we
leave it to the reader. We do not focus on this case to avoid
discussion of existence and positivity of Green's function with
these boundary conditions and uniqueness of solution to boundary
value problem.
\end{remark}
\begin{remark}
A comparison of two simple particular cases of the Proposition leads
to a curious observation. Take the entire upper boundary with
homogeneous Dirichlet or homogeneous Neumann boundary conditions.
In both cases the drift $D_t$
vanishes. These two Gaussian free fields with mutually singular laws are
therefore both coupled with dipolar SLE${}_4$.
\end{remark}

\section{Couplings in doubly connected domains}
\label{sec: doubly connected}
In this section we address the question of couplings of SLE and GFF in
doubly connected domains. We first consider punctured disc and exhibit a
coupling of GFF with radial SLE, and then consider annuli $\bA_p$.
The non-simply connectedness requires in many cases non trivial monodromies
of the free field --- in order to obtain couplings with single valued fields
we need to compactify the field, that is consider free field with values on
a circle.
In physics literature considerations of lattice model height functions in multiply
connected domains or in the presence of vortices, and considerations of
operator algebra and modular invariance of conformal field theories have
both lead to the study of such compactified free fields.

Throughout this chapter, $x$, $x_1$, $\dots$ denote points on the boundary of
an annulus $\bA_r$ for some $r>0$, and the derivatives $\pder{x}$, $\pder{x_1}$,
$\dots$ will be
taken in counterclockwise direction, both for inner and outer boundary.

\subsection{Compactified GFF and radial SLE${}_4$}
\label{sec: radial SLE}
We first investigate the solutions to basic equations (\ref{eq: M-eq ci}) and
(\ref{eq: C-eq ci}) in the radial case. We will see that the solution to these
equation will not be a harmonic function in the disc, but rather a harmonic
function with monodromy. This situation has also been considered in
\cite{Dubedat-SLE_and_free_field}.

We use the Loewner chain (\ref{eq: Loewner D}) to describe the growth process in
$\domain_0 = \bD$, and for radial SLE${}_4$ we have the driving process
$X_t = \exp( \ii 2 B_t )$ so in the absence of other marked points but the
tip of the growth the basic equations
(\ref{eq: M-eq ci}) and (\ref{eq: C-eq ci}) read
\begin{align}
\label{MainEqRad1}
& \im \left\{ 2 \; \pderS{x} F + z \frac{x+z}{x-z} \; \pder{z} F \right\} \; = \; 0
\qquad \textrm{ and } \\
\label{MainEqRad2}
& \der{t} C_{\domain_t}(z_1,z_2) \; = \;  - 4 \; \im \Big( \pder{x} F(z_{1};x) \Big) \;
\im \Big( \pder{x} F(z_{2};x) \Big) .
\end{align}
As usually, with $C_{\domain_t}$ the Dirichlet Green's function,
Hadamard's formula suggests the solution to (\ref{MainEqRad2}),
with the ambiguity of a sign and an additive real constant. Namely we have
\[ \pder{x} F(z;x) \; = \; \pm 2 \ii \lambda \; S_x(z) + \const \; = \;
\pm \ii \; \frac{\lambda}{\pi} \; \frac{x+z}{x-z} + \const , \]
expressed in terms of the Schwarz kernel $S_x(z)$ in the unit disc.
The sign of $\pder{x} F$ is unimportant, but the constant
will have to vanish.
We warn the reader that here, for the first time, it is important not to
confuse the derivatives $\pder{x}$ w.r.t. the length parameter
on the boundary with derivatives w.r.t. the position of the marked
point $x$.

The function $F$ can be taken invariant under rotations, and we get
\begin{equation}
\label{rotinv}
\ii \, z \, \pder{z} F + \pder{x} F \; = \; 0 .
\end{equation}
Using this in \eqnref{MainEqRad1}, and integrating explicitly
gives
$$
F(z;x) \; = \; \frac{\lambda}{\pi} \big(( 2\log (x-z) - \log (z) \big) 
$$
and 
$M(z;x) = \frac{\lambda}{\pi} \big( - \arg(z) + 2 \, \arg (x-z) \big)$.
The function $M$ is not single-valued. However, all the formulas
we have used make sense: as soon as we fix the branch of $M(z)$, the
branch of $M(g_t(z))$ will also be fixed by continuity. We can thus
define a multi-valued harmonic function $M(z)$ in the punctured disc
$\bD \setminus \{0\}$, such that it has monodromy of
$2 \lambda = \sqrt{\frac{\pi}{2}}$ around zero, and the boundary conditions
have a jump of $2\lambda$ at the point $x$, being otherwise locally
constant. Adding this function to a zero Dirichlet boundary valued GFF
in $\bD$, one obtaines a multi-valued GFF $\Phi$ of the same monodromy,
which could be interpreted as a single valued free field with values in
$\bR / 2 \lambda \bZ$.

\begin{remark}
\thmref{thm: coupling} is not directly formulated for multivalued free fields,
but this problem is superficial. It is easy to see that the corresponding
single valued free field on the universal cover of $\bD \setminus \set{0}$
(with periodic covariance, in particular) is coupled with the growth
process obtained by lifting the radial SLE${}_4$ to the universal cover.
Here and in the sequel we nevertheless prefer to talk about either
multivalued free field or free field with values on a circle
$\bR / 2 \lambda \bZ$.
\end{remark}

\subsection{Compactified GFF and standard annulus SLE${}_4$}
A natural generalization of the radial SLE to annuli of finite modulus
is the standard annulus SLE. We will now show that at $\kappa=4$ it is
coupled with a multivalued free field having Neumann boundary conditions
on the inner boundary component of the annulus and jump-Dirichlet boundary
conditions on the outer boundary component.

The starting domain is taken to be $\domain_0 = \bA_p$, and we use
the Loewner chain (\ref{eq: Loewner A}) to describe the growth process.
The conformal maps $g_t : \bA_p \setminus \hull_t \rightarrow \bA_{p-t}$
uniformize the complements of the hull to thinner annuli, so even with
strict conformal invariance we have to specify the mean and the covariance
for all annuli $\bA_{p-t}$. We take the covariance $C_t = C_{\bA_{p-t}}$
to be the Green's function with Dirichlet boundary conditions on
$\set{|z|=1}$ and Neumann boundary conditions on $\set{|z|=e^{-p+t}}$.
The mean $M_t = M_{\bA_{p-t}}$ will be represented as the imaginary part
of a multivalued analytic function $F_t$ defined on $\bA_{p-t}$.
Correspondingly, the equation (\ref{eq: M-eq ci}) should be generalized
to
\begin{equation}
\label{MainEq1Annulus}
\im \left\{ \frac{\kappa}{2} \; \pderS{x} F_t(z;x) + V^{p-t}_x(z) \; \pder{z} F_t(z;x)
    + \pder{t} F_t(z;x) \right\} \; = \; 0 .
\end{equation}
Recall that $V^{p-t}_x(z) = 2\pi \, z \, S_{x}^{p-t}(z)$ where $S_{x}^{p-t}(z)$
is the Schwarz kernel in the annulus $\bA_{p-t}$, as specified in
\secref{sec: Loewner chains}.

The equation (\ref{eq: C-eq ci}) is exactly the same as before
and Hadamard's formula applies, so we find $\im (\pder{x} F_t(z;x))$ to
be equal to a multiple of the Poisson kernel in the annulus $\bA_{p-t}$
with zero Dirichlet boundary conditions on the outer boundary and Neumann
boundary conditions on the inner one. We can write
$ \pder{x} F_t(z;x) = 2 \ii \lambda \; \tilde{S}_x^{p-t}(z)$
where $\tilde{S}$ is the Schwarz kernel with the following boundary
conditions: $\re (\tilde{S}_x(z)) = \delta_x(z)$ on $\{|z|=1\}$ and
$\im (\tilde{S})=0$ on $|z|=e^{t-p}$.
As in the radial case, we have rotational invariance (\ref{rotinv})
which allows us to rewrite (\ref{MainEq1Annulus}) as 
\begin{align}
\label{MainEq1AnnulusBis}
& \im (\fvanish) \; \equiv \; 0 , \qquad \textrm{ where } \\
\nonumber
& \fvanish \; := \; 2\ii \; \pder{x} \tilde{S}_x^{p-t}(z)
    - 2\pi \; S_x^{p-t}(z) \; \tilde{S}_x^{p-t}(z) + \frac{1}{2\lambda}
    \; \partial_t F_t(z;x)
\end{align}
We now prove that $\im(\fvanish)$ is a harmonic function
in the annulus satisfying 
\begin{itemize}
\item $\im (\fvanish) = 0$ on the outer part of the boundary
\item $\nder \, \im (\fvanish) = 0$ on the inner part of the boundary
\item $\im (\fvanish)$ is bounded.
\end{itemize}
This will imply the equation (\ref{MainEq1AnnulusBis}), and consequently
establish (\ref{M-eq}) and (\ref{C-eq}). First two
boundary conditions for $\im(H)$ obviously hold on
$\bdry \bA_{p-t} \setminus \set{x}$:
if $M_{\domain_t}$ satisfies those conditions for all $t$, then
so does its drift $\im (\fvanish)$.
So, we only need to prove that $\im (\fvanish)$ has no singularity
at $x$. Without loss of generality, assume $t=0$. The expansions of the
two Schwarz kernels at $z = x$ coincide up to constant order
\begin{align*}
S_x(z) \; = \; \frac{-x}{\pi (z-x)} - \frac{1}{2\pi}  + \Ord{z-x}
\qquad \textrm{ and } \qquad 
\tilde{S}_x(z) \; = \; \frac{-x}{\pi (z-x)} - \frac{1}{2\pi}  + \Ord{z-x} ,
\end{align*}
as follows from the condition that their real parts give the delta function
on the outer boundary.
Plugging these into (\ref{MainEq1AnnulusBis}) shows that the possible
singularities at $x$ cancel out. 

We summarize the result of this subsection in the following proposition:
\begin{proposition}
\label{prop: AnnulusNeumann}
For any $p>0$, let $M^p(z;x)$ be the unique multi-valued harmonic function
in the annulus $\bA_p$ satisfying the following properties:
\begin{itemize}
\item $M^p(z;x)$ obeys zero Neumann boundary conditions on the inner
boundary circle $\{|z|=e^{-p}\}$;
\item $M^p(z;x)$ has a jump-Dirichlet boundary conditions on the
outer boundary circle, namely, for any branch of $M^p(z;x)$ there exist
$n\in \bZ$ such that
$M^p(x e^{\pm i\theta};x) \equiv \mp \lambda + 2\lambda n =
\mp\sqrt{\frac{\pi}{8}}+2\sqrt{\frac{\pi}{8}} n$ for small
positive $\theta$, and  $M^p(x,z)$ is locally constant on
$\{|z|=1\} \setminus \{x\}$.
\end{itemize}
As the free field $\Phi$ in $\bA_p$, $p>0$, take the sum
$\Phi(z) = \Phi_0(z) + M^p(z;x)$, where $\Phi_0$ is a GFF in $\bA_p$ with
zero Neumann boundary conditions on $\{|z|=e^{-p}\}$ and zero Dirichlet boundary
conditions on $\{|z|=1\}$. In other domains define the free field in the
same manner, using conformal invariance.
Then the standard annulus SLE${}_4$ is coupled with these free fields in
the sense of \thmref{thm: coupling}.
\end{proposition}

\subsection{More marked points on the outer boundary}
In this subsection, we extend the result above to the case
of additional marked points $x_1,x_2,\dots$ on the outer boundary of the
annulus $\bA_p$.
The free field will have locally constant Dirichlet boundary conditions
with jumps $2\lambda=\sqrt{\frac{\pi}{2}},2\lambda_1,2\lambda_2,\dots$ at
$x,x_1,x_2,\dots$ . If we impose zero Neumann boundary conditions on the
inner boundary, then, for any choice of $(\lambda_j)$ we find a variant of
SLE${}_4$ which is coupled with this field. If jumps add up to zero, one
can also impose Dirichlet boundary condition on the inner boundary. In all
cases, drifts of driving processes are computed explicitly.

We start with the case of one additional marked point and Neumann boundary
conditions on the inner boundary. Let $\tilde{S}_x(z)$ be as in the previous
section.
\begin{proposition}
\label{prop: AnnulusNeumannMore}
For any $p>0$, let $M^p(z;x,x_1)$ be the 
multi-valued harmonic function in the annulus $\bA_p$ satisfying
the following properties:
\begin{itemize}
\item $M^p(z;x,x_1)$ obeys zero Neumann boundary conditions on the inner
boundary $\{|z|=e^{-p}\}$;
\item $M^p(z;x,x_1)$ has a jump-Dirichlet boundary conditions on the outer
part of the boundary with jumps $-2\lambda=-\sqrt{\frac{\pi}{2}}$ at $x$
and $-2\lambda_1$ at $x_1$.
\end{itemize}
Let $\Phi_0$ be a GFF in $\bA_p$ with zero Neumann boundary conditions
on $\{|z|=e^{-p}\}$ and zero Dirichlet boundary conditions on $\{|z|=1\}$.
In $\bA_p$, $p>0$, take the GFF as
$\Phi(z) = M^p(x,z) + \Phi_0(z)$, and for other domains use conformal
invariance. These free fields are coupled with an annulus SLE${}_4$ variant
defined using (\ref{eq: Loewner A}) with the driving process
\[ \ud X_t \; = \; \ud W_{4t}
    - \ii \pi \rho \; \tilde{S}^{p-t}_{g_t(x_1)}(X_t) \; \tau_{X_t} \; \ud t ,
\]
where $W$ stands for the Brownian motion on $\{|z|=1\}$ and
$\rho = 2 \frac{\lambda_1}{\lambda}$.
\end{proposition}
\begin{remark}
The letter $\rho$ is used here analogously to the case of ordinary
SLE${}_\kappa(\rho)$. Indeed, the drift of the driving process
of SLE${}_\kappa(\rho)$ in $\bH$ is $\frac{\rho}{X_t - g_t(x_1)}
= -\ii \pi \rho \, S^{\bH}_{g_t(x_1)}(X_t)$, where $S^\bH_x(z)$ is the Schwarz
kernel in $\bH$ with Dirichlet boundary conditions.
The value $\rho = 2 \frac{\lambda_1}{\lambda}$ is
also what one gets in the case of simply connected domains and piecewise constant
Dirichlet boundary conditions with jump of size $2 \lambda_1$ at a marked point.
\end{remark}
\begin{proof}
The proof essentially repeats the one of \propref{prop: AnnulusNeumann}.
Let us stress the differences. The first basic equation
(\ref{eq: M-eq ci}) now reads 
\begin{equation}
\label{MainEq1AnMore}
\im \left\{ \frac{\kappa}{2} \pderS{x} F + V_x(z) \pder{z} F + \pder{t} F
    + \frac{D_t}{\ii x} \; \pder{x} F
    - 2\pi \ii S_x^{p-t}(x_1) \; \pder{x_1} F \right\} \; = \; 0 ,
\end{equation}
whereas the second one (\ref{eq: C-eq ci}) is exactly the same as in
\propref{prop: AnnulusNeumann}. Hence we should choose
$\pder{x} F_t(z;x) = 2 \ii \lambda \, \tilde{S}_{x,p-t}(z)$ with the same
$\tilde{S}$ (note that this identity holds true for the choice of $M$
made in the assertion). The rotational invariance (\ref{rotinv}) now reads
$$ \ii z \pder{z} F + \pder{x} F +\pder{x_1} F = 0
$$
and we rewrite (\ref{MainEq1AnMore}) as  
\begin{align}
\label{MainEq1AnMoreBis}
\im (\fvanish) \; = \; & 0 \qquad \text{, where} \\
\nonumber
\fvanish \; := \; & 2\ii \; \pder{x} \tilde{S}^{p-t}_x(z)
    - 2\pi \; S^{p-t}_x(z) \; \tilde{S}^{p-t}_x(z) \\
\nonumber
& + \frac{1}{2 \lambda} \big(
    2\pi \ii \; S^{p-t}_x(z) \; \pder{x_1} F + \pder{t} F_t + D_t \; \pder{x} F
    - 2\pi \ii \; S_x^{p-t}(x_1) \; \pder{x_1}F \big) .
\end{align}
As before, it suffices to show that for an appropiate choice of $D_t$
the function $\fvanish$ is bounded and has zero imaginary part on the
outer part of the boundary and constant real part on the inner one.
The boundary conditions on $\bdry \bA_{p-t} \setminus \{x,x_1\}$ follow
immediately. A first-order singularity at $x_1$ might be produced by
the two terms containing $\pder{x_1}F$, but we see that their
contributions exactly cancel each other. As we have seen in the proof
of \propref{prop: AnnulusNeumann},
$2\ii \, \pder{x} \tilde{S}^{p-t}_x(z) - 2\pi \, S^{p-t}_x(z) \, \tilde{S}^{p-t}_x(z)$
is bounded near $x$, and hence there could only be a first-order singularity
produced by $2\pi \ii \, S_x^{p-t}(z) \pder{x_1} F + D_t \pder{x} F$. The
choice of $D_t$ made in the assertion is exactly to guarantee that it
vanishes.
\end{proof}

We now consider the case of Dirichlet boundary conditions on the inner
boundary circle.
\begin{proposition}
\label{prop: AnnulusDirichletMore}
For any $p>0$, let $M^p(z;x,x_1)$ be the unique harmonic function in the
annulus $\bA_p$ satisfying the following boundary conditions:
\begin{itemize}
\item $M^p(z;x,x_1)=\lambda=\sqrt{\frac{\pi}{8}}$ on counterclockwise arc from
$x_1$ to $x$ and 
\item $M^p(z;x,x_1)=-\lambda$ on counterclockwise arc from $x$ to $x_1$
\item $M^p(z;x,x_1)=\mu \in \bR$ on the inner boundary circle $\{|z|=e^{-p}\}$
\end{itemize}
Let $\Phi_0(z)$ be a GFF in $A_p$ with zero Dirichlet boundary conditions.
Then the GFF $M^p(x,x_1,z)+\Phi_0(z)$, transported to other domains using conformal
invariance, is coupled in the sense of \thmref{thm: coupling}
with the following annulus SLE${}_4$ variant.
The driving process $X_t$ in (\ref{eq: Loewner A}) is given by
\begin{align*} \ud X_t \; = \; \ud W_{4t} 
    + D_t \, \tau_{X_t} \; \ud t ,
\end{align*}
where $W_t$ stands for the Brownian motion on $\{|z|=1\}$, and the drift is
explicitly
\begin{align*}
D_t \; = \; - \ii \pi \rho \; S^{p-t}_{g_t(x_1)}(X_t)
    + \frac{2\pi}{p-t} \left( \frac{\mu}{2 \lambda} +
            \frac{\length{[X_t,g_t(x_1)]} - \pi}{2 \pi} \right)
\end{align*}
with $\rho = -2$ and $\length{[x,x_1]}$ denoting
the length of the counterclockwise boundary arc from $x$ to $x_1$. 
\end{proposition}
\begin{remark}
Since $\rho = -2 = \kappa - 6$, it is easy to show using coordinate changes
of the kind described in \cite{SW-coordinate_changes},
that in the limit $p \rightarrow \infty$ one recovers a chordal SLE${}_4$
in $\bD$ from $x$ to $x_1$. This limit therefore
degenerates to the basic example of Schramm \& Sheffield discussed in
\secref{sec: SSExample}.
\end{remark}
\begin{remark}
The annulus SLE with the above driving process was proposed in
\cite{HBB-free_field_in_annulus}, based on considerations of regularized
free field partition function with these boundary conditions. That article also
computes the probabilities that the curve passes to the left or right of the
inner boundary circle, and finds that there is a non-zero probability for the
curve to touch the inner circle only if $-\lambda < \mu < \lambda$ ---
as anticipated for a discontinuity line of the free field between the
levels $\pm \lambda$.
\end{remark}
\begin{proof}
For the above choice of $M$, if $F$ is holomorphic function
such that $M = \im (F)$, we have that $\im ( \pder{x} F )$ is equal to
Dirichlet boundary valued Poisson kernel
$P_x^{p-t}(z) = \re \left( S^{p-t}_x(z) + \frac{1}{2 \pi (p-t)} \log(z) \right)$,
exactly as required by  (\ref{eq: C-eq ci}) and Hadamard's formula.
Observe that the harmonic conjugate of $M$ is not a single-valued function and one
should be careful defining $\re (F)$.
A rotationally invariant definition of $F$ is given by the following formula:
\[ F_t(z;x,x_1) \; := \; - \lambda \ii \; \int_x^{x_1} S_w^{p-t}(z) \, |dw|
    + \lambda \ii \; \int_{x_1}^x S_w^{p-t}(z) \, |dw|
    + \frac{ \ii \; \log(z/x) }{(t-p)}
        \left( \mu- 2\lambda \frac{\pi-\length{[x,x_1]}}{2\pi} \right) , \]
the intergals being along the boundary in a counterclockwise direction.
We have the following expressions for derivatives of $F$:
\begin{align}
\label{partxF}
\pder{x} F \; = \; & 2 \lambda \ii \; \left( S_x^{p-t}(z) + R_1 \right)
\\
\label{partx1F}
\pder{x_1} F \; = \; & 2\lambda \ii \; \left( - S_{x_1}^{p-t}(z) + R_2 \right),
    \qquad \text{ where } \\
\nonumber
R_1(x,x_1,z,t) \; = \; & - \ii \; \frac{\frac{\mu}{2\lambda}}{t-p}
    + \ii \; \frac{\pi-\length{[x,x_1]}}{2\pi(t-p)} - \frac{\log (z/x)}{2 \pi (t-p)} \\
\nonumber
R_2(x,x_1,z,t) \; = \; & \frac{\log\frac{z}{x}}{2\pi(t-p)}
\end{align}
As in the proof of \propref{prop: AnnulusNeumannMore} we write
the first basic equation (\ref{M-eq}) using the above expressions and
rotational invariance as
\begin{align}
\label{MainEq1AnMoreBis1}
\im ( \fvanish) \; = \; & 0 ,\qquad \text{ where } \\
\nonumber
\fvanish \; := \; & 2 \ii \; \pder{x} (S_x^{p-t}(z)+R_1)
    - 2 \pi S_x^{p-t}(z) (S_x^{p-t}(z) + R_1) \\
\nonumber
& + 2 \pi \big( S_x^{p-t}(x_1) - S_x^{p-t}(z) \big)
    \big( -S_{x_1}^{p-t}(z) + R_2 \big)
+ \frac{1}{2 \lambda} \Big( \ii D_t \; (S_x^{p-t}+R_1) + \pder{t} F \Big)
\end{align}
Now the function $\fvanish$ is possibly multi-valued.
To prove (\ref{MainEq1AnMoreBis1}) we check that $\fvanish$ satisfies
the following properties:
\begin{itemize}
\item $\im (\fvanish) = 0$ on $\bdry \bA_{p-t} \setminus \{x,x_1\}$;
\item Any branch of $\im (\fvanish)$ is bounded near $x$ and $x_1$. 
\end{itemize}
Since $\fvanish$ clearly cannot grow faster than linearly at infinity on
the universal cover, these conditions guarantee (\ref{MainEq1AnMoreBis1}).
The first condition is justified as in the previous propositions.
The singularities at $x_1$ clearly cancel out. It remains to
handle possible singularities at $x$. As before, 
$2 \ii \, \pder{x} (S_x^{p-t}(z)+R_1) - 2\pi \, S_x^{p-t}(z) \, S_x^{p-t}(z)$
is bounded near $x$, and three terms that have
first-order singularities at $x$ remain in $\fvanish$:
\[ -2\pi \; S_x^{p-t}(z) \, R_1
+ 2\pi \; S_x^{p-t}(z) \, \left( S_{x_1}^{p-t} - R_2 \right)
+ \ii \, D_t \, S_x^{p-t}(z). \]
We see that since $R_1+R_2$ doesn't contain $\log z$, the choice
$\ii \, D_t = 2\pi(-S_{x_1}(x)+ (R_1 + R_2) |_{z=x})$ guarantees vanishing of the
singularity for all branches of $\fvanish$, and we are done.
\end{proof}
\begin{remark}
The extension of \propDref{prop: AnnulusNeumannMore}{prop: AnnulusDirichletMore}
to the case of several marked point $x_1,x_2,\dots$ with jumps
$2\lambda_1,2\lambda_2,\dots$ is straightforward; proofs are literally the
same. The drift term $D_t$ is just the sum
$D_t = \sum_j \frac{\lambda_j}{\lambda} D^{j}_t$ where $D^{j}_t(x,x_j)$
is the drift we would have if we only had one jump of size $2\lambda_j$
at $x_j$. Indeed, one may observe that procedure of determining the drift
term for $F$ is in fact linear in $F$. One should remember, however, that
in the Dirichlet case the construction only makes sense if all jumps
add up to $0$, including the jump of size
$2 \lambda = \sqrt{\frac{\pi}{2}}$ at $x$.
\end{remark}

\subsection{Compactified GFF with a marked point on the inner boundary}
In this section we consider the case when an additional marked point
$x_1$ is on the inner part of the boundary. The mean of the field $M$
will be a multi-valued harmonic function obeying Dirichlet boundary
conditions with jumps $-2\lambda$ both at $x$ and $x_1$. However, these
conditions do not define $M$ completely, we should also define the
change of fixed branch of the function $M$ along the radius, say, from
$e^{-p}$ to $1$.

Let $\hat{M}^p(z)$ be the unique multi-valued harmonic
function in $\bA_p$ such that any continuous branch in
any sector
$\{ r e^{\ii \theta} \, : \, e^{-p} < r < 1 , \;
\theta_1 < \theta < \theta_2 \}$ determined by angles
$\theta_2 \in [0,2\pi[$ and $\theta_1 \in ]\theta_2-2\pi,\theta_2[$,
has boundary
values $\lambda \, (\sign(\theta) + 2 n)$ at $z = e^{\ii \theta}$ and
$z = e^{-p+\ii \theta}$.
The function $M_t(z;x,x_1)$ in $\bA_{p-t}$ is constructed by
continuously moving the discontinuity points of boundary conditions of 
$\hat{M}^{p-t}$ from $1$ to $x$ and from $e^{-p}$ to $x_1$. Hence, $M$ is in fact
a multi-valued harmonic function in $z$ that depends on $x$, $x_1$, $t$
and the choice of $\arg x_1 - \arg x$. More precisely, represent $\hat{M}^p$ as the
imaginary part of a multivalued analytic function $\hat{F}^p$, and $M_t$ as the
imaginary part of
\begin{align*}
F_t(z;\arg x,\arg x_1) \; = \; & \hat{F}^{p-t}(z) +
 2 \lambda \ii \; \int_0^{\arg x} S_{e^{\ii \theta}}^{p-t}(z) \; \ud \theta
+ 2 \lambda \ii \; \int_0^{\arg x_1} S^{\inv; p-t}_{e^{\ii \theta+t-p}}(z) \; \ud \theta \\
& \qquad -2 \lambda \ii \; \frac{ \log \frac{z}{x}}{ 2 \pi (t-p)} \; \arg x
-2 \lambda \ii \; \frac{t-p - \log \frac{z}{x}}{2 \pi (t-p)} \; \arg x_1 .
\end{align*}
Here $S^{\inv;p}_{y}(z):=S^p_{e^{-p}/y}(e^{-p}/z)$. With this definition, the
function $F$ is invariant under rotations. We will sometimes write it as
function of $x$ and $x_1$ where the branch of the argument will be clear from the
context. We have the following proposition:
\begin{proposition}
\label{prop: AnnulusJump}
Let $M$ be as above, and let $\Phi_0(z)$ be a GFF in $\bA_p$ with zero Dirichlet
boundary conditions. Consider the multi-valued GFF defined in $\bA_p$ as
$M^p(x,x_1,z)+\Phi_0(z)$ and in other domains by conformal invariance.
It is coupled in the sense of \thmref{thm: coupling}
with the annulus SLE${}_4$ variant whose driving process $X_t$ in
(\ref{eq: Loewner A}) satisfies
\begin{align*}
\ud X_t \; = \; & \ud W_{4t} + D_t \, \tau_{X_t} \; \ud t \qquad \textrm{ with} \\
D_t \; = \; & - 2 \pi \ii \left( S^{\inv; p-t}_{g_t(x_1)}(X_t) - \frac{1}{2\pi} \right)
    + \frac{\arg g_t(x_1)-\arg X_t}{p-t} .
\end{align*}
\end{proposition}
\begin{proof}
The proof literally repeats the one of \propref{prop: AnnulusDirichletMore}.
We now have, as in \eqnDref{partxF}{partx1F},
\begin{align*}
\pder{x} F \; = \; & 2 \lambda \ii \; \left( S_x^{p-t}(z)+R_1 \right) \\
e^{-p} \, \pder{x_1} F \; = \; &
    2\lambda \ii \; \left( S^{\inv; p-t}_{x_1}(z) + R_2 \right)
\qquad \textrm{, where}  \\
R_1(x,x_1,z,t) \; = \; & -\frac{\log\frac{z}{x}}{2\pi(t-p)}
    + \ii \; \frac{(\arg x - \arg x_1) }{2 \pi (t-p)} \\
R_2(x,z,t) \; = \; & - \frac{1}{2\pi}\left(1-\frac{\log\frac{z}{x}}{t-p}\right) .
\end{align*}
and we find that the basic equations are verified provided that
\[ \ii \, D_t \, = \, 2\pi \left( S^{\inv}_{x_1}(x) + (R_1+R_2)|_{z=x} \right) . \]
\end{proof}
\begin{remark}
One can write explicitly the stochastic differential equation satisfied
by the process $\arg(g_t(x_1)) - \arg(X_t)$. It turns out to be a 
Brownian bridge which at time $t=p$ hits $0$. Therefore, at $t=p$, the curve
hits $x_1$ with a winding determined by the initial choice of
$\arg(x_1) - \arg(x)$.
\end{remark}

\subsection{Generalizations to $\kappa \neq 4$ for Dirichlet boundary conditions}
\label{sec: Dirichlet general kappa}
In the case of Dirichlet boundary conditions, one can generalize the
previous couplings to $\kappa\neq 4$. As the example of the section
(\ref{sec: SSExample}) shows, the rule that associates a field to a
domain is not conformally invariant: if we have a conformal map
$\varphi:\domain_1\rightarrow \domain_2$, then 
\begin{equation}
\label{eq: preSchwarzian}
M_{\domain_1;x_1,x_2,\dots}(z) \; = \;
M_{\domain_2;\varphi(x_1),\varphi(x_2),\dots}(\varphi(z))+ \alpha_\kappa \; \arg \varphi'(z),
\end{equation}
where $\alpha_\kappa = \frac{4-\kappa}{2 \sqrt{2\pi\kappa}}$ as in
\eqnref{eq: ad hoc term 3}.
The covariance, however, is still the Dirichlet Green's function. 
Consider the annulus $\bA_p$ with two marked points $x\in\{z:|z|=1\}$, $x_1\in \partial \bA_p$, and let $M^{p}_4(z,x,x_1)$ be one of the functions $M^{p}$ defined in proposition (\ref{prop: AnnulusDirichletMore}) or (\ref{prop: AnnulusJump}). We define
\[ M^{p}_{\kappa}(z,x,x_1) \; := \;
\sqrt{\frac{4}{\kappa}}M^{p}_{4}(z,x,x_1) - 
\alpha_\kappa \; \arg z.
\]
This is a multi-valued harmonic function (with a single-valued derivative);
the monodromy is equal to
$\big( \kappa - 6 \big) \lambda_\kappa $.
For an arbitrary doubly-connected domain, we define the mean of the
field by conformal map to an annulus and the rule
(\ref{eq: preSchwarzian}); in particular, for
$\bA_p\backslash \hull_t$ we have
\begin{equation}
\label{eq: MeanNot4}
M^{\bA_p\backslash \hull_t}_{\kappa}(z,x,x_1)\; = \;
\sqrt{\frac{4}{\kappa}} M^{p-t}_{4}
\big( g_t(z);g_t(x),g_t(x_1) \big) + 
\alpha_\kappa \; \Big( \arg g'_t(z)-\arg g_t(z) \Big).
\end{equation}

We have the following proposition:
\begin{proposition}
A GFF defined as above (with marked point on the outer or
inner boundary) is coupled with annulus SLE defined using
(\ref{eq: Loewner A}) with the driving process
\[ \ud X_t \; = \; \ud W_{\kappa t} + D_t \, \tau_{X_t} \; \ud t, \]
$D_t$ being the same as in \propref{prop: AnnulusDirichletMore}
or \ref{prop: AnnulusJump} correspondingly.
\end{proposition}
\begin{proof}

The additional term
$\alpha_\kappa \, (\arg g'_t(z)-\arg g_t(z))$
has finite variation, hence the proof of \mbox{(\ref{eq: C-eq ci})} will
be the same as before (we have adjusted the coefficient in front of
$M$ to compensate the change of speed for $W_{\kappa t}$).
Note, however, that without that term the proof of
\propref{prop: AnnulusDirichletMore} (correspondingly \ref{prop: AnnulusJump})
would fail for $\kappa\neq 4$ because the coefficient in front
of the first term of the definition of $\tilde{F}$ (see the
equation (\ref{MainEq1AnMoreBis1})) changes from $2$ to
$\frac{\kappa}{2}$, hence the second-order singularities at
$x$ would not cancel out anymore. We now show that the additional
term exactly compensates this effect, without destroying zero
Dirichlet boundary conditions elsewhere.

Simple geometric considerations show that
$\ud (\arg g'_t(z)-\arg g_t(z))=0$ when
$g_t(z)\in\partial \bA_{p-t}\backslash \set{X_t}$. One has
\begin{align*}
\partial_t \log g'_t(z) \; = \; &
V'_{X_t}(g_t(z)) \; = \; 2\pi g_t(z)S'_{X_t}(g_t(z))+2\pi S_{X_t}(z)
\qquad \textrm{ and }\\
\partial_t \log g_t(z) \; = \; & 2\pi S_{X_t}(z) .
\end{align*}
Recall the rotational invariance of the Schwarz kernel:
$\partial_x S_x(z)+ \ii z \, S'_x(z)=0$, and the fact that the second-order
singularity of $\fvanish$ comes from its first term
$\frac{\kappa}{2} \ii \partial_{x}S_{x,p-t}(z)$.
Comparing the coefficients finishes the proof.
\end{proof}


%

\appendix

\section{Non-commutation at $\kappa \neq 4$ for general boundary conditions}
\label{sec: commutation}
This appendix discusses a difference between Dirichlet boundary conditions
and other boundary conditions concerning the couplings with SLEs at $\kappa \neq 4$.
In the case of Schramm \& Sheffield treated in \secref{sec: SSExample} as well as
those of \secref{sec: Dirichlet general kappa}
we have remarked that for the coupling with SLE variants
with $\kappa \neq 4$, it suffices to modify the the the boundary
conditions of the one point function $M$ by a harmonic interpolation of the
winding of the boundary. In other cases no such claims were made,
and we now explain why these cases indeed don't admit a generalization of this
sort.

For the sake of concreteness we detail the argument only in the simplest case
of combined jump-Dirichlet and Riemann-Hilbert boundary conditions as treated in
\secref{sec: strip RH}.
Recall that $\pder{x} F$ is determined by (\ref{eq: C-eq ci})
and the Hadamard formula. One then defines, as in (\ref{eq: ad hoc term}),
\[ M_{\domain_t}(z) = \im \left( F(g_t(z);X_t) + E_t(z) \right) , \]
which contains a process of finite variation $(E_t(z))_{t \geq 0}$ introduced
in order to restore the martingale property  of the mean (\ref{eq: M-eq ci})
at the cost of relaxing strict conformal invariance. Concretely,
\begin{align} \label{eq: ad hoc general}
E_t(z) = \int_0^t \im \Big( J_{X_s}(g_s(z)) \Big) \, \ud s ,
\end{align}
where $J_x(z)$ is a multiple of the derivative of the appropriate
Schwarz kernel, see \eqnDref{eq: ad hoc term 2}{eq: ad hoc strip}.
A question naturally arises:
is the modified formula for $M_{\domain_t}$ consistent with having a
function $M_{\domain;x,x_1, \ldots, x_n}$
associated to any domain with marked points? Does (\ref{eq: ad hoc general})
depend on the full history $(g_s)_{s \in [0,t]}$ of the Loewner chain, or
can it be expressed as a function of domain $\domain_t$ only, as is the
case in (\ref{eq: ad hoc term 3})? 

Imagine two different Loewner chains that in the end uniformize the same
hull. The prototype is a hull $K = \hull_- \cup \hull_+$ consisting of two small
pieces $\hull_+$, $\hull_-$ away from each other, located roughly at
$\xi_+, \xi_- \in \bdry \domain_0$. We can uniformize $K$ by first
uniformizing one piece and then what remains of the other. Suppose
that the local half plane capacities of $\hull_+$ and $\hull_-$ are
$\eps_+$ and $\eps_-$, respectively. In the calculations below we keep track
of terms of order $\eps_\pm$ as well as the second order cross terms of type
$\eps_+ \eps_-$, but we omit other second order and higher order terms.
Write the uniformizing maps of complements of $\hull_\pm$
constructed by a Loewner chain (\ref{eq: Loewner}) as
\begin{align*}
g_\pm \; : \; & \domain_0 \setminus \hull_\pm \rightarrow \domain_0 \\
g_\pm(z) \; \approx \; &
    z \; + \; \eps_\pm \, V_{\xi_\pm}(z) + \cdots .
\end{align*}
After having thus removed one piece $\hull_\pm$, we are left with the hull
$\wtil{K}_\mp = g_\pm(\hull_\mp)$ whose local half plane capacity is
\[ \wtil{\eps}_\mp \; \approx \;  \eps_\mp \; |(g_\pm)'(\xi_\mp)|^2 + \cdots
\; \approx \; \eps_\mp + 2 \eps_\pm \eps_\mp \; (V_{\xi_\pm})'(\xi_\mp) + \cdots \]
and the hull $\wtil{K}_\mp$ can be uniformized by a map constructed by the
same Loewner fields
\begin{align*}
\wtil{g}_\mp \; : \; & \domain_0 \setminus \wtil{K}_\mp \rightarrow \domain_0 \\
\wtil{g}_\mp(z) \; \approx \; &
    z \; + \; \wtil{\eps}_\mp \, V_{\wtil{\xi}_\mp}(z) + \cdots ,
\end{align*}
where $\wtil{\xi}_\mp$ is the location of the hull $\wtil{K}_\mp$
\[ \wtil{\xi}_\mp = g_\pm(\xi_\mp) \approx
    \xi_\mp + \eps_\pm \, V_{\xi_\pm}(\xi^\mp) + \cdots . \]
We then have two conformal maps
\[ \wtil{g}_+ \circ g_- \; \textrm{ and } \; \wtil{g}_- \circ g_+
\quad : \quad \domain_0 \setminus K \rightarrow \domain_0 . \]
In practise the Loewner vector fields are chosen to be the unique
ones preserving some normalization condition, so the two maps must actually
be equal. In any case, we can ask whether formula (\ref{eq: ad hoc general})
gives the same answer for the hull $K$ built in the two possible ways.
The two expressions for $E_t$ are approximately
\[ \eps_\mp \; J_{\xi_\mp}(z) + \wtil{\eps}_\pm \; J_{\wtil{\xi}_\pm} (g_\mp(z)) , \]
so their difference can be expressed expanding in all small parameters
\begin{align}
\nonumber
\Delta E_t \; \approx \; \eps_+ \eps_-
\Big\{ & 2 \, (V_{\xi_-})'(\xi_+) \, J_{\xi_+}(z)
        - 2 \, (V_{\xi_+})'(\xi_-) \, J_{\xi_-}(z) \\
\nonumber
& + V_{\xi_-}(\xi_+) \,\pder{x} J_{\xi_+}(z)
        - V_{\xi_+}(\xi_-) \,\pder{x} J_{\xi_-}(z) \\
\label{eq: commutation}
& + V_{\xi_-}(z) \, \pder{z} J_{\xi_+}(z)
        - V_{\xi_+}(z) \, \pder{z} J_{\xi_-}(z)
\Big\} + \cdots
\end{align}
For $E_t$ to be a function of the hull $K$ only, and not of the history of the
Loewner chain, it is necessary that $J$ satisfies the functional
equation that makes the above expression vanish identically.

As is already clear from considerations of the chordal SLE${}_\kappa$ coupling,
in particular \eqnref{eq: ad hoc term 
3}, the function
$J_x(z) = \const \; \frac{1}{(z-x)^2}$ satisfies the appropriate equation with
$V_x(z) = \frac{2}{z-x}$ chosen according to the Loewner flow
(\ref{eq: Loewner}).

In the strip $\bS$ we considered jump-Dirichlet boundary conditions on $\bR$
and Riemann-Hilbert on $\bR + \ii \pi$. We chose correspondingly
$J_x(z) = \const \; \pder{x} \tilde{S}_x(z)$, where $\tilde{S}_x(z)$ is the
Schwarz kernel (\ref{eq: explicit RH Schwarz}) with the same boundary
conditions. A direct computation shows that
with the appropriate Loewner vector field
$V_{x}(z) = \coth(\frac{z-x}{2})$, this $J_x(z)$ produces a non vanishing
difference in (\ref{eq: commutation}). It is therefore not possible to
generalize the coupling of \secref{sec: strip RH} to $\kappa \neq 4$ in the
manner analogous to Dirichlet boundary conditions.


\section{Local half-plane capacity and \propref{prop: hloc}}
\label{sec: Loewner lemma}
Most of the statements of \propref{prop: hloc} are
standard Loewner chains techniques (and may be found in the literature for
all particular cases we deal with in this paper), so we leave the proof to
the reader. We will only discuss the slightly less standard statement about
the local half-plane capacity. 

Let $\domain$ be a planar domain, $x\in\partial \domain$, and let
$\partial \domain$ be analytic in a neigborhood of $x$. Let $(\hull_t)$
be a family of growing compact hulls in $\overline{\domain}$,
$\lim\limits_{t\rightarrow 0} \hull_t=\{x\}$.  Henceforth we assume
that $x=0$, the tangent to the boundary at $x$ is parallel to the
real line, and that the inner normal at $0$ points to the upper half-plane. 

Let $\Psi$ be a harmonic function in $\domain\backslash \hull_t$
with the following boundary conditions:
\begin{itemize}
\item $\Psi(z)=\dist(z,\partial \domain)$ on $\partial \hull_t$
\item $\Psi(z)=0$ on $\partial \domain\backslash \hull_t$
\end{itemize}

Let $r>0$ be small enough, so that $\domain\cap\{|z|=r\}$ consists of
one arc $\{r e^{\ii \theta} : \theta_1<\theta<\theta_2\}$.
If the diameter of $\hull_t$ does not exceed $r$, define
\begin{equation}
\label{lhcap}
L^{\domain}_{\hull_t,r}=\frac{1}{\pi}\int_{\theta_1}^{\theta_2}
    \Psi (r e^{\ii\theta}) r \sin (\theta) \; \ud \theta.
\end{equation}

If $\domain=\mathbb{H}$, then $L^{\domain}_{\hull_t,r}$ is well-known
to be the half-plane capacity of $\hull_t$. We will thus call this
quantity the \textit{local half-plane capacity at distance $r$}.

It is easy to see that $L^{\domain}_{\hull_t,r}$ satisfies the following
two properties, that express its stability under slight changes of the domain:
\begin{itemize}
\item Let $\phi:\domain_1\rightarrow \domain_2$ be a conformal map
such that $\phi(0)=0$ and $\phi'(0)=1$. Then
$|\frac{L^{\domain_1}_{\hull_t,r}}{L^{\domain_2}_{\hull_t,r}}-1|\leq C r$.
\item Let $R>r$, and
$\domain_1\cap B_R(0)=\domain_2\cap B_R(0)$.
Then $|\frac{L^{\domain_1}_{\hull_t,r}}{L^{\domain_2}_{\hull_t,r}}-1|
\leq C \frac{r}{R}$.
\end{itemize}

These properties allow us to define
$\partial_t \lochcap(\hull_t)|_{t=0} :=
\lim\limits_{r\rightarrow 0}\partial_t L^{\domain}_{\hull_t,r}$.
It remains unchanged under conformal maps $\phi$ as in the first property
above, and is equal to the derivative of the half-plane capacity of
$\hull_t$ if $\partial \domain$ coincides with the real line in some
neighborhood of zero. Henceforth we assume without loss of generality
that this is the case.

Now, let $\hull_t$ be generated by a Loewner chain as in 
\propref{prop: hloc}. We first claim that, when computing
$\partial_t \lochcap(\hull_t)|_{t=0}$, we can replace $\Psi (z)$ by
$\im z -\im g_t(z)$ in the integral (\ref{lhcap}). Indeed, the
difference $H(z):= \Psi (z)-\im z +\im g_t(z)$ is a harmonic function;
$H(z)\equiv 0$ on $\partial \domain\cap B_R(0)$ for some constant
$R$, and $|H(z)|\leq C t$ elsewhere on $\partial \domain$. Hence
$|H(r e^{\ii\theta})|\leq C\frac{r}{R} t$, and this is negligible
when we take $r$ to zero. 
 
However, we have
\[ \partial_t \, \im g_t(z)|_{t=0} \; = \; \im (V_0(z)) \; = \;
\im(\frac{2}{z}) + O(1) \; = \;\partial_t \, \im h_t(z)|_{t=0} +O(1),
\qquad r\rightarrow 0, \]
where $h_t(z)$ is the conformal map from $\mathbb{H}\backslash \hull_t$
to $\mathbb{H}$ (i. e. the solution to the half-plane Loewner equation).
Since in the half-plane the formula (\ref{lhcap}) defines the
half-plane capacity, we are done.

\bigskip

\noindent
{\bf Acknowledgements:}
Work supported by Swiss National Science Foundation and
\mbox{ERC AG CONFRA}.

\def\cprime{$'$} \def\cprime{$'$} \def\cprime{$'$}


\begin{thebibliography}{Zha04b}

\bibitem[BB03]{BB-zig_zag}
M.~Bauer and D.~Bernard,
\newblock {SLE}, {CFT} and zig-zag probabilities,
\newblock in \textsl{ Proceedings of the conference `Conformal Invariance and
  Random Spatial Processes', Edinburgh}, 2003.

\bibitem[BB06]{BB-2d_growth_processes}
M.~Bauer and D.~Bernard, \textsl{ 2{D} growth processes: {SLE} and {L}oewner
  chains},
\newblock Phys. Rep. \textbf{ 432}(3-4), 115--222 (2006),
  {[arXiv:math-ph/0602049]}.

\bibitem[BBC09]{BBC-near_critical}
M.~Bauer, D.~Bernard and L.~Cantini, \textsl{ Off-critical SLE(2) and SLE(4): a
  field theory approach},
\newblock J. Stat. Mech. , P07037 (2009), {[arXiv:0903.1023]}.

\bibitem[BBH05]{BBH-dipolar}
M.~Bauer, D.~Bernard and J.~Houdayer, \textsl{ Dipolar stochastic {L}oewner
  evolutions},
\newblock J. Stat. Mech. (3), P03001, 18 pp. (electronic) (2005).

\bibitem[Car04]{Cardy-SLE_kappa_rho}
J.~Cardy,
\newblock {SLE}(kappa,rho) and Conformal Field Theory,
\newblock 2004.

\bibitem[Dub07]{Dubedat-SLE_and_free_field}
J.~Dub{\'e}dat,
\newblock {SLE} and the free field: {P}artition functions and couplings,
\newblock [arXiv:0712.3018], 2007.

\bibitem[HBB10]{HBB-free_field_in_annulus}
C.~Hagendorf, M.~Bauer and D.~Bernard,
\newblock The Gaussian free field and SLE(4) on doubly connected domains,
\newblock 2010.

\bibitem[KM]{KM-free_fields}
N.-G. Kang and N.~Makarov,
\newblock in preparation.

\bibitem[Kyt06]{Kytola-SLE_kappa_rho}
K.~Kyt{\"o}l{\"a}, \textsl{ On conformal field theory of {SLE}(kappa, rho)},
\newblock J. Stat. Phys. \textbf{ 123}(6), 1169--1181 (2006),
  {[arXiv:math-ph/0504057]}.

\bibitem[Law05]{Lawler-conformally_invariant_processes}
G.~F. Lawler,
\newblock \textsl{ Conformally invariant processes in the plane}, volume 114 of
  \textsl{ Mathematical Surveys and Monographs},
\newblock American Mathematical Society, Providence, RI, 2005.

\bibitem[MS09]{MS-massive_SLEs_ICMP}
N.~Makarov and S.~Smirnov,
\newblock Off-critical lattice models and massive SLEs,
\newblock Proceedings of ICMP, to appear., 2009.

\bibitem[MZ]{MZ-free_fields}
N.~Makarov and D.~Zhan,
\newblock in preparation.

\bibitem[SS05]{SS-harmonic_explorer}
O.~Schramm and S.~Sheffield, \textsl{ Harmonic explorer and its convergence to
  {${\rm SLE}\sb 4$}},
\newblock Ann. Probab. \textbf{ 33}(6), 2127--2148 (2005),
  {[arXiv:math.PR/0310210]}.

\bibitem[SS09]{SS-contour_lines}
O.~Schramm and S.~Sheffield, \textsl{ Contour lines of the two-dimensional
  discrete Gaussian free field},
\newblock Acta Math. \textbf{ 202}, 21--137 (2009), {[arXiv:math.PR/0605337]}.

\bibitem[SW05]{SW-coordinate_changes}
O.~Schramm and D.~B. Wilson, \textsl{ S{LE} coordinate changes},
\newblock New York J. Math. \textbf{ 11}, 659--669 (electronic) (2005),
  {[arXiv:math.PR/0505368]}.

\bibitem[Wer04]{Werner-random_planar_curves}
W.~Werner,
\newblock Random planar curves and {S}chramm-{L}oewner evolutions,
\newblock in \textsl{ Lectures on probability theory and statistics}, volume
  1840 of \textsl{ Lecture Notes in Math.}, pages 107--195, Springer, Berlin,
  2004.

\bibitem[Zha04a]{Zhan-thesis}
D.~Zhan,
\newblock \textsl{ Random Loewner Chains in Riemann Surfaces},
\newblock PhD thesis, California Institute of Technology, 2004.

\bibitem[Zha04b]{Zhan-SLE_in_doubly_connected_domains}
D.~Zhan, \textsl{ Stochastic {L}oewner evolution in doubly connected domains},
\newblock Probab. Th. Rel. Fields \textbf{ 129}(3), 340--380 (2004),
  {[arXiv:math/0310350]}.

\end{thebibliography}
\end{document}